\documentclass[10pt,a4paper,twocolumn]{article}

\usepackage[T1]{fontenc}    
\usepackage{times}          
\usepackage[utf8]{inputenc}

\usepackage{authblk}        
\usepackage{etoolbox}       
\usepackage[margin=2cm]{geometry}

\usepackage[numbers]{natbib}
\usepackage{amsmath,amssymb,mathtools}
\usepackage{siunitx}
\usepackage{graphicx}
\usepackage{booktabs}
\usepackage{hyperref}
\usepackage[nameinlink]{cleveref}
\usepackage{microtype}
\usepackage{xcolor}
\usepackage{subcaption}
\usepackage{bm}
\usepackage{makecell}
\usepackage{tikz}
\usepackage{tabularx}
\usepackage[outline]{contour}
\usepackage{standalone} 
\usetikzlibrary{arrows.meta, calc, backgrounds, shapes.geometric}
\usepackage{pifont}

\newcommand{\cmark}{\ding{51}}
\newcommand{\xmark}{\ding{55}}

\newtheorem{remark}{Remark}

\newcommand{\body}{{\mathcal{B}}}
\newcommand{\world}{{\mathcal{W}}}
\DeclareSIUnit{\rad}{rad}

\def\tsc#1{\csdef{#1}{\textsc{\lowercase{#1}}\xspace}}
\tsc{WGM}
\tsc{QE}

\newcommand{\samethanks}{\protect\footnotemark[1]}

\title{\LARGE \bf Nonlinear System Identification Nano-drone Benchmark}

\author[1]{Riccardo Busetto\thanks{Equal contribution. Corresponding author: {\tt\small name.surname@supsi.ch}}}
\author[1]{Elia Cereda\samethanks}
\author[1]{Marco Forgione}
\author[1]{Gabriele Maroni}
\author[1]{Dario Piga}
\author[1,2]{Daniele Palossi}

\affil[1]{Dalle Molle Institute for Artificial Intelligence (IDSIA), USI-SUPSI, Lugano, Switzerland {\tt\small {name}.{surname}@supsi.ch}}
\affil[2]{Integrated Systems Laboratory (IIS), ETH Z\"urich, Switzerland\hspace{5cm} 
{\tt\small dpalossi@iis.ee.ethz.ch}}

\date{} 

\begin{document}

\maketitle 

\begin{abstract}
We introduce a benchmark for system identification based on \SI{75}{k} real-world samples from the \emph{Crazyflie~2.1 Brushless} nano-quadrotor, a sub-\SI{50}{\gram} aerial vehicle widely adopted in robotics research.
The platform presents a challenging testbed due to its multi-input, multi-output nature, open-loop instability, and nonlinear dynamics under agile maneuvers.
The dataset comprises four aggressive trajectories with synchronized \emph{4-dimensional motor inputs} and \emph{13-dimensional output measurements}.
To enable fair comparison of identification methods, the benchmark includes a suite of \emph{multi-horizon prediction metrics} for evaluating both one-step and multi-step error propagation.
In addition to the data, we provide a detailed description of the platform and experimental setup, as well as baseline models highlighting the challenge of accurate prediction under real-world noise and actuation nonlinearities.
All data, scripts, and reference implementations are released as open-source at \url{https://github.com/idsia-robotics/nanodrone-sysid-benchmark} to facilitate transparent comparison of algorithms and support research on agile, miniaturized aerial robotics.
\par\vspace{1em}
\end{abstract}

\section{Introduction} \label{sec:introduction}

Standardized benchmarks play a central role in accelerating scientific progress by enabling fair comparison, reproducibility, and rapid testing of new algorithms. 
Their impact has been particularly evident in machine learning and deep learning, where well-defined datasets and evaluation protocols have catalyzed major breakthroughs~\citep{deng2009imagenet, bellemare2013arcade, wu2018moleculenet}.
In the field of system identification, the \textit{Nonlinear Benchmark Initiative}~\citep{champneys2024baseline, schoukens2017three}\footnote{available at \url{https://www.nonlinearbenchmark.org/}} serves a similar purpose by promoting reproducible and comparable research efforts, highlighting the importance of high-quality datasets and reference methods for rigorous and comparable evaluation.

However, within the domain of aerial robotics, the availability of such standardized benchmarks remains limited.
While public datasets exist, they typically focus on perception-centric tasks such as visual-inertial odometry or Simultaneous Localization and Mapping (SLAM)~\citep{thalagala2024mun, Delmerico19icra, burri2016euroc}, rather than dynamical modeling or system identification.
Even when high-quality flight data is available from aggressive flight sequences as in~\cite{bauersfeld2021neurobem}, it is typically provided as a resource to support a specific method, rather than as a structured benchmark with unified evaluation metrics and diverse baselines. 
This leaves a gap for researchers who wish to rapidly test and compare new identification algorithms without the burden of designing their own evaluation protocols or reference implementations.

This gap is critical for miniaturized platforms, such as \textit{nano-drones} (quadrotors with a mass below \SI{50}{\gram} and a diameter below \SI{10}{\centi\meter}), whose adoption has skyrocketed due to their low cost and suitability for safe indoor operation (\cite{lamberti2024simtoreal, bonato2023ultra}).
Yet nano-drones operate in a uniquely challenging dynamical regime: their extremely low mass and compact geometry limit the thrust and torque generated by their motors and propellers.
Consequently, they are more sensitive to disturbances and less capable of rapid corrective action, making them significantly more challenging to model and control during agile flight compared to standard-sized vehicles.
Therefore, this emerging robotic platform represents a compelling candidate for our novel system identification benchmark.

Despite these challenges and the growing scientific interest in such platforms, there is currently no publicly available benchmark that provides real flight data for nano-drones, together with a well-defined evaluation protocol suitable for system identification and control.
Existing datasets for small aerial vehicles either offer limited scope, target different tasks, or lack the structure required for systematic algorithmic comparison~\citep{szecsi2024deep}.
As a result, researchers must often collect their own trajectory data, which is typically restricted to simple maneuvers or narrow experimental conditions, thereby preventing reproducibility and slowing progress.

In this work, we introduce a benchmark specifically designed to address this gap.
The benchmark is built around the \emph{Crazyflie~2.1 Brushless} nano-quadrotor, a commercially available, open-source hardware and software/firmware platform that has recently gained traction in robotics and control research~\citep{akbari2024tiny,cereda2024trainonthefly,lamberti2024simtoreal}. 
Despite its small size, the Crazyflie exhibits complex non-linear dynamics, as it is a \textit{multi-input multi-output} system that is inherently \textit{open-loop unstable}, displaying \textit{strongly nonlinear} behavior when operated in aggressive conditions.
These characteristics make it an appealing and challenging testbed for nonlinear system identification and model-based control.

The primary contribution of this work is a \emph{ready-to-use, high-quality benchmark} that, to the best of our knowledge, constitutes the first publicly available dataset of real flight trajectories collected on a nano-drone. 
Importantly, the dataset not only covers favorable or quasi-static motion, but also includes aggressive maneuvers that challenge the dynamics of such a small platform and expose the difficulty of achieving accurate multi-step prediction on highly agile nano-scale vehicles.
Specifically, the benchmark provides:
\begin{enumerate}
    \item \textbf{A thoroughly-collected dataset of real flight trajectories} recorded using synchronized onboard sensing and motion-capture ground truth, specifically designed to excite the system across a broad dynamic range. The dataset allows for an \textit{as is} use, without requiring access to costly hardware or motion-capture infrastructure.
    \item \textbf{A fully specified evaluation protocol} based on multi-step-ahead prediction, enabling consistent and transparent comparison of identification methods.
    \item \textbf{Reference baseline models}, including a nominal physics model, a hybrid physics--learning architecture, and purely data-driven models, illustrating the intrinsic difficulty of the task and providing starting points for future research.
    \item \textbf{Open-source data and code}, lowering the barrier to entry for system identification and control studies on agile, miniaturized aerial platforms.
    \item \textbf{A calibrated motor-speed mapping}, extending the off-the-shelf drone's firmware with the bidirectional DSHOT protocol for accurate telemetry of the motor speeds (RPM). 
\end{enumerate}

The remainder of the paper is organized as follows.
Section~\ref{sec:related} surveys the relevant literature, identifies the main shortcomings of existing datasets and identification studies, and further motivates the contributions presented in this work.
Section~\ref{sec:settings} provides a detailed description of quadrotor operation, introducing the notation, conventions, and the fundamental dynamic model. 
Section~\ref{sec:setup} presents the specific Crazyflie platform and the experimental setup employed for data acquisition. 
In Section~\ref{sec:experiments}, we describe in detail the executed trajectories and the resulting dataset, including preprocessing and the proposed evaluation strategy for this benchmark. 
Section~\ref{sec:results} introduces several baseline models, ranging from a purely physics-based formulation to black-box approaches, for performing $N$-step-ahead predictions.
Finally, Section~\ref{sec:conclusions} concludes the paper by highlighting the main challenges and outlining potential directions for future research.

\begin{figure}
    \centering \includegraphics[width=\linewidth, trim=0 20 0 30, clip]{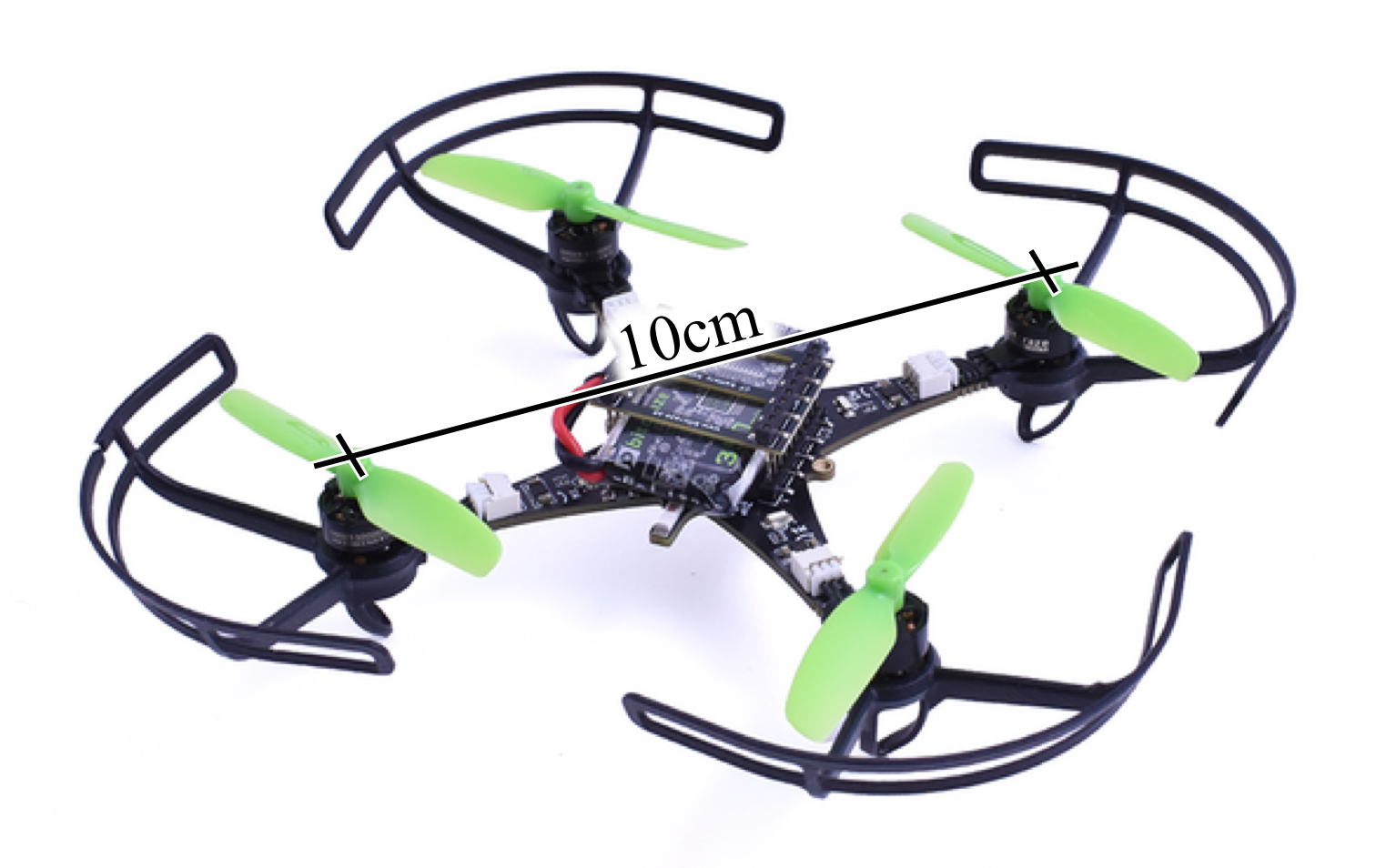}
    \caption{Crazyflie 2.1 Brushless nano-drone.}
    \label{fig:crazyflie}
\end{figure}
\section{Related Work} \label{sec:related}

\begin{table}[t]
\centering
\caption{Aerial drones taxonomy by vehicle class-size.}
\label{tab:taxonomy}
\resizebox{\linewidth}{!}{%
\begin{tabular}{lcccc p{0.9cm} p{1.3cm} p{1.3cm} p{1.2cm} p{0.7cm}}
\toprule
\small
Vehicle class & Diameter & Weight & Power & Device\\
\midrule
\text{\textit{standard-size}} & $\geq$ \SI{50}{\centi\meter} & $\geq$ \SI{1}{\kilo\gram}	& $\geq$ \SI{100}{\watt} & GPU\\
\text{\textit{micro-size}} & $\sim$ \SI{25}{\centi\meter} & $\sim$ \SI{500}{\gram} & $\sim$ \SI{50}{\watt} & CPU\\
\text{\textit{nano-size}} & $\sim$ \SI{10}{\centi\meter} & $\sim$ \SI{50}{\gram} & $\sim$ \SI{5}{\watt} & MCU\\
\text{\textit{pico-size}} & $\leq$ \SI{2}{\centi\meter} & $\leq$ \SI{5}{\gram} & $\leq$ \SI{0.1}{\watt} & ULP\\
\bottomrule
\end{tabular}
}
\end{table}

Research on quadrotor system identification encompasses a wide range of platforms, modeling objectives, and data collection methodologies.
In Table~\ref{tab:taxonomy}, we report a taxonomy on aerial vehicles, commonly used in the robotics community.
Our taxonomy reports the diameter, weight, total power consumption (both motors and electronics), and onboard processing device of the platforms.
The class of devices available onboard is a direct consequence of the total power envelope available, as only $\sim$15\% of this total can be spent on the electronics (both sensing and computing), see~\cite{Wood2017}.

Standard-sized drones have a diameter of more than \SI{50}{\centi\meter}, a total power envelope of more than \SI{100}{\watt}, and a payload that allows them to afford powerful CPUs and GPUs~\citep{niculescu2022fly}.
Micro-sized drones are slightly smaller in size, but with a lower total power envelope of approximately \SI{50}{\watt}, they can still host powerful embedded CPUs, such as NVIDIA Jetson/Xavier devices.
Nano-sized drones, with their sub-\SI{50}{\gram} weight and a few 100s \SI{}{\milli\watt} for onboard processing, can only afford resource-constrained MicroController Units (MCUs), featuring a few 100s \SI{}{\kilo\byte} on-chip memories.
Finally, the smaller class of vehicles is represented by pico-sized UAVs, which are as small as a coin but remain out of reach for widespread adoption~\citep{Wood2017} and would require Ultra-Low Power (ULP) chips, consuming only a few tens of \SI{}{\milli\watt}.

In this context, the majority of publicly available datasets have been developed either for standard-sized multirotors or for perception-focused tasks, with only limited attention given to dynamical modeling of nano-scale vehicles. 
This section reviews the most relevant efforts and highlights the absence of benchmarks tailored to nano-quadrotor system identification.

\paragraph{Aggressiveness of trajectories}
In the aerial robotics literature, a trajectory is typically considered as \emph{aggressive} when it drives the vehicle through fast translational or rotational motions relative to its actuation 
capabilities. 
Concretely, aggressive flight is characterized by (i) translational speeds exceeding
$2$--$3\,\mathrm{m/s}$ for nano-scale platforms (and $>10\,\mathrm{m/s}$ for larger multirotors), (ii) angular rates above $400$--$600^\circ/\mathrm{s}$, or (iii) linear accelerations approaching or exceeding $1$--$2$ times gravity.
Such trajectories induce actuator saturation, strong axis couplings, and aerodynamic effects far from near-hover conditions, all of which are essential for evaluating nonlinear system-identification algorithms.

\subsection{System Identification on Larger Quadrotors}

Several high-fidelity identification studies target custom or commercial multirotors equipped with specialized sensing. \citet{sun2019quadrotor} perform gray-box identification of a standard-sized quadrotor using high-speed flight data recorded in the Open Jet Facility wind tunnel, revealing strong aerodynamic interactions at high speed.
\citet{bauersfeld2021neurobem}~propose the NeuroBEM hybrid aerodynamic model for a custom quadrotor, using aggressive free-flight trajectories collected inside a large
25$\times$25$\times$\SI{8}{\meter} motion-capture arena.
While these datasets provide rich aerodynamic information, the platforms are significantly larger than nano-quadrotors and rely on expensive, non-standard infrastructure.

The DronePropA dataset~\citep{ismail2025dronepropa} focuses on fault detection in a standard-sized \SI{1.5}{\kilo\gram} quadrotor and includes 130 flight sequences with various propeller defects.
Although extensive, it targets health monitoring rather than predictive modeling, and the platform size places it far from the dynamical regime of sub-\SI{50}{\gram} nano-scale vehicles.

\subsection{Nano-Quadrotor Identification}

Only a limited number of works consider nano-scale platforms.
\citet{szecsi2024deep} study deep-learning-based identification of a Crazyflie~2.1 using simulation data, without releasing real-flight measurements or standardized evaluation protocols.
\citet{forster2015crazyflie} provides an early characterization of the Crazyflie 2.0, identifying inertial parameters, thrust curves, and motor dynamics using separate static and dynamic tests.
Their work provides valuable physical insights, but (i) does not include aggressive flight trajectories, (ii) does not release a public dataset, and (iii) does not provide an evaluation protocol for multi-step-ahead prediction.
Beyond these studies, several perception-oriented datasets (e.g., VIO/SLAM-focused micro-UAV datasets) include nano-scale platforms.
Still, they do not provide the state, actuation, and ground-truth signals required for system identification or control design.

\subsection{Positioning of the Contribution}

Existing resources suffer from at least one of the following limitations: (i) platforms are medium- or large-scale~\citep{sun2019quadrotor, bauersfeld2021neurobem, ismail2025dronepropa}; (ii) datasets target perception or fault detection rather than dynamical modeling; (iii) data are not publicly available, simulation-only, or not suitable for multi-step prediction; or (iv) no unified evaluation protocol is given.

Table \ref{table:comparison} provides a consolidated comparison of existing datasets and system-identification studies on quadrotors, highlighting their limitations in terms of platform size, data availability, aggressiveness of flight trajectories, and suitability as benchmarks.
This comparison further motivates the need for a dedicated nano-drone benchmark, as introduced in this work. 
Indeed, to the best of our knowledge, no publicly available benchmark provides real, aggressive flight data for a \emph{nano-quadrotor} together with a standardized system-identification protocol and baseline models.
The present work fills this gap by releasing a complete benchmark tailored for control and system identification.

\begin{table*}[t]
\centering
\caption{Comparison of related datasets and system-identification studies on quadrotors.}
\label{table:comparison}
\renewcommand{\arraystretch}{1.3}
\resizebox{\linewidth}{!}{%
\begin{tabular}{lcccccc p{2.1cm} p{1.6cm} p{2.8cm} p{1.5cm} p{1.7cm} p{1.8cm}}
\toprule
\textbf{Work} &
\textbf{Class-size} &
\textbf{Real/Sim} &
\textbf{\makecell{Application}} &
\textbf{Open-source} &
\textbf{Benchmark} &
\textbf{Aggressive} \\
\midrule
\cite{sun2019quadrotor} & Micro & Real & Aerodynamics Gray-box SysID & \xmark & \cmark & \xmark \\
\cite{bauersfeld2021neurobem} & Standard & Real/Sim & Aerodynamics / Hybrid modeling & \cmark & \cmark & \cmark \\
\cite{ismail2025dronepropa} & Standard & Real & Fault diagnosis & \cmark & \xmark & \xmark \\
\cite{szecsi2024deep} & Nano & Sim & Learning-based SysID & \xmark & \cmark & \xmark \\
\cite{forster2015crazyflie} & Nano & Real & Parameter estimation & \xmark & \xmark & \xmark \\
\midrule
\textbf{This work} & \textbf{Nano} & \textbf{Real} & \textbf{SysID \& control} & \textbf{\cmark} & \textbf{\cmark} & \textbf{\cmark} \\
\bottomrule
\end{tabular}}
\end{table*}
\section{Quadrotor Model} \label{sec:settings}

\subsection{Notation}

Throughout this paper, scalars are denoted by regular letters $s, S$, vectors by lowercase bold symbols $\mathbf{v}$, and matrices by uppercase bold symbols $\mathbf{M}$.
The quadrotor dynamics are formulated using two right-handed reference frames: the inertial \emph{world} frame ${\mathcal{W}}$, with axes $\{\mathbf{x}_{\world}, \mathbf{y}_{\world}, \mathbf{z}_{\world}\}$, fixed to the ground and oriented such that $\mathbf{z}_{\world}$ points upward; and the \emph{body} frame ${\body}$, attached to the vehicle, with origin at the center of mass, and equipped with axes $\{\mathbf{x}_{\body}, \mathbf{y}_{\body}, \mathbf{z}_{\body}\}$ oriented as shown in Fig.~\ref{fig:crazyflie-convention}.
The orientation of $\body$ with respect to $\world$ is represented by a unit Hamilton quaternion $\mathbf{q} \in \mathbb{H}$ following the scalar-last convention, i.e., $\mathbf{q} = [q_x,\, q_y,\, q_z,\, q_w]^\top$ with $\|\mathbf{q}\| = 1$.
The set of all unit quaternions forms the three-dimensional sphere
$\mathbb{S}^3 = \{ \mathbf{q} \in \mathbb{H} \mid \|\mathbf{q}\| = 1 \}$.
Each unit quaternion $\mathbf{q} \in \mathbb{S}^3$ is associated with a
rotation matrix $\mathbf{R}(\mathbf{q}) \in \mathrm{SO}(3)$ through a
smooth mapping $\mathbf{R} : \mathbb{S}^3 \to \mathrm{SO}(3)$.
Finally, the operator $\otimes$ denotes quaternion multiplication.

Unless otherwise specified, vectors are expressed in world coordinates; e.g., $\mathbf{p}$ denotes the position of the drone's center of mass with respect to ${\world}$ expressed in world coordinates. 
To stress that a vector is expressed in a particular reference frame, we attach the frame name as a subscript. 
For example, $\mathbf{f}_\body$ denotes the forces acting on the drone expressed in the {body} frame.

\subsection{Quadrotor and propeller dynamics} \label{sec:quad_motor_dynamics}

The state $\mathbf{x} \in \mathbb{R}^{13}$ of the quadrotor is defined as:
\begin{equation}
    \mathbf{x} = 
    \begin{bmatrix} 
        \mathbf{p}^\top\ \mathbf{v}^\top\ \mathbf{q}^\top\ \bm{\omega}^\top
    \end{bmatrix}^\top,
\end{equation}
where $\mathbf{p} = [p_x,\, p_y,\, p_z]^\top \in \mathbb{R}^3$ [m] is the position of the vehicle’s center of mass expressed in the world frame~$\world$, and $\mathbf{v} = [v_x,\, v_y,\, v_z]^\top \in \mathbb{R}^3$ [m/s] is the corresponding linear velocity, also expressed in $\world$.
The unit quaternion $\mathbf{q} = [q_x,\, q_y,\, q_z,\, q_w]^\top$ represents the orientation of the body frame~$\body$ with respect to the world frame~$\world$.
Finally, $\bm{\omega} = [\omega_x,\, \omega_y,\, \omega_z]^\top \in \mathbb{R}^3$ [rad/s] denotes the angular velocity of the body relative to the world, expressed in body coordinates~$\body$. 
The components $\omega_x$, $\omega_y$, and $\omega_z$ correspond to rotation rates around the body axes $\mathbf{x}_\body$, $\mathbf{y}_\body$, and $\mathbf{z}_\body$, respectively.

The continuous-time dynamics of the quadrotor can be written compactly as:
\begin{equation}
\dot{\mathbf{x}} =
\begin{bmatrix}
    \dot{\mathbf{p}} \\[3pt]
    \dot{\mathbf{v}} \\[3pt]
    \dot{\mathbf{q}} \\[3pt]
    \dot{\bm{\omega}}
\end{bmatrix}
=
\begin{bmatrix}
    \mathbf{v} \\[3pt]
    \tfrac{1}{m}\mathbf{R}(\mathbf{q})\,\mathbf{f}_\body + \mathbf{g} \\[3pt]
    \tfrac{1}{2}\,\mathbf{q} \otimes \begin{bmatrix} \bm{\omega} \\[2pt] 0 \end{bmatrix} \\[8pt]
    \mathbf{J}^{-1}\!\left( \bm{\tau} - \bm{\omega} \times (\mathbf{J}\bm{\omega}) \right)
\end{bmatrix},
\label{eq:quad_dynamics}
\end{equation}
where $m$ is the quadrotor mass, $\mathbf{J}$ is the inertia matrix in body coordinates, $\mathbf{f}_\mathcal{B} = [0,\, 0,\, T]^\top$ is the total force acting on the drone expressed in the body frame, $\bm{\tau} = [\tau_x,\, \tau_y,\, \tau_z]^\top$ is the control torque vector, and $\mathbf{g} = [0,\, 0,\,-g]^\top$ is the gravity vector in world-frame coordinates, with $g = 9.81~\mathrm{m/s^2}$.

For control design, the input is typically defined as:
\begin{equation}
    \label{eq:control_inputs}
    \mathbf{u}_{\mathrm{ctrl}} = 
    \begin{bmatrix} T \ \tau_x \ \tau_y \ \tau_z \end{bmatrix}^\top,
\end{equation}
where $T$ is the total thrust acting along the body $z$-axis $\mathbf{z}_{\body}$, and 
$(\tau_x, \tau_y, \tau_z)$ are the torques about the body axes $\mathbf{x}_{\body}$, $\mathbf{y}_{\body}$, and $\mathbf{z}_{\body}$, corresponding respectively to \emph{roll}, \emph{pitch}, and \emph{yaw} rotations.

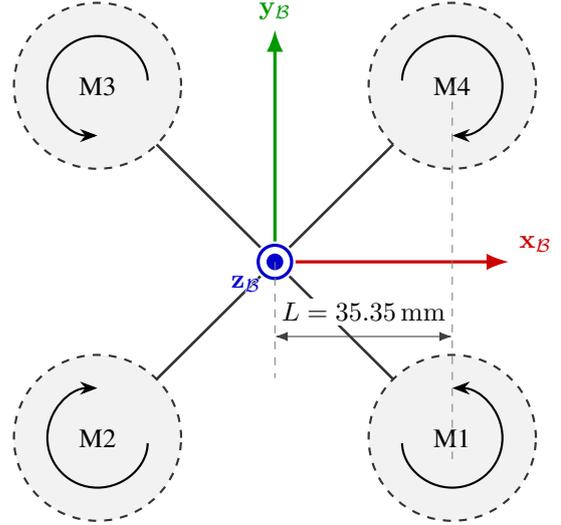
\begin{figure}
    \centering
    \begin{tikzpicture}[
    scale=2.2,
    >=Latex, 
    axis/.style={->, very thick, line cap=round},
    motor/.style={circle, draw=black!80, dashed, fill=gray!10, thick, minimum size=2.2cm, inner sep=0pt},
    arm/.style={line width=1pt, black!80, line cap=round}, 
    dim/.style={thin, <->, darkgray},
    rot_arrow/.style={thick, ->, >=Stealth, shorten >=0pt, shorten <=2pt},
    label_bg/.style={fill=white, inner sep=1.5pt, text=black}
]

    \def\armradius{1.5} 
    \def\arcrad{0.3} 

    \coordinate (Center) at (0,0);
    \coordinate (M1_Pos) at (315:\armradius);
    \coordinate (M2_Pos) at (225:\armradius);
    \coordinate (M3_Pos) at (135:\armradius);
    \coordinate (M4_Pos) at (45:\armradius);

    \begin{scope}[on background layer]
        \draw[arm] (Center) -- (M1_Pos);
        \draw[arm] (Center) -- (M2_Pos);
        \draw[arm] (Center) -- (M3_Pos);
        \draw[arm] (Center) -- (M4_Pos);
    \end{scope}
    
    
    \node[motor] (M1) at (M1_Pos) {M1};
    \draw[rot_arrow] ($(M1) + (180:\arcrad)$) arc (180:450:\arcrad); 
    
    \node[motor] (M2) at (M2_Pos) {M2};
    \draw[rot_arrow] ($(M2) + (0:\arcrad)$) arc (0:-270:\arcrad);

    \node[motor] (M3) at (M3_Pos) {M3};
    \draw[rot_arrow] ($(M3) + (0:\arcrad)$) arc (0:270:\arcrad); 

    \node[motor] (M4) at (M4_Pos) {M4};
    \draw[rot_arrow] ($(M4) + (180:\arcrad)$) arc (180:-90:\arcrad); 

    \draw[axis, green!60!black] (0,0) -- (0, 1.4) node[anchor=south] {\contour{white}{$\mathbf{y}_{\mathcal{B}}$}};
    \draw[axis, red!80!black] (0,0) -- (1.4, 0) node[anchor=south west] {\contour{white}{$\mathbf{x}_{\mathcal{B}}$}};

    \filldraw[white] (Center) circle (0.12); 
    \filldraw[fill=white, draw=blue!80!black, very thick] (Center) circle (0.10); 
    \fill[blue!80!black] (Center) circle (0.05); 
    \node[anchor=north east, inner sep=2pt] at (-0.05,-0.05) {\contour{white}{\color{blue!80!black}$\mathbf{z}_{\mathcal{B}}$}};

    \draw[dashed, gray] (Center) -- (0, -0.7); 
    \draw[dashed, gray] (M4_Pos) -- ++(0, -2.25); 
    
    \coordinate (DimStart) at (0, -0.3);
    \coordinate (DimEnd)   at ({1.5*cos(45)}, -0.3);
    
    \node[label_bg] (DimLabel) at ($(DimStart)!0.5!(DimEnd)$) {$L = \SI{35.35}{\milli\meter}$};
    
    \draw[dim] ($(DimStart)+(0,-0.15)$) -- ($(DimEnd)+(0,-0.15)$);

\end{tikzpicture}%
    \vspace{0.25cm}
    \caption{``$\times$'' configuration and axis convention adopted in this work (geometry based on \cite{giernacki2017crazyflie}).}
    \label{fig:crazyflie-convention}
\end{figure}

In this work, however, we take as input the individual propeller angular velocities (in rad/s):
\begin{equation}
    \label{eq:propeller_inputs}
    \mathbf{u} = 
    \begin{bmatrix}
        \Omega_1 \ \Omega_2 \ \Omega_3 \ \Omega_4
    \end{bmatrix}^\top,
\end{equation}
since these are the physical quantities that generate the above-mentioned thrust and torques.
Assuming a quadratic relation between the propellers’ angular velocities and the resulting forces/torques~\citep{furrer2016rotors}, and considering the ``$\times$'' quadrotor configuration characterizing the Crazyflie (see Figure~\ref{fig:crazyflie-convention}), the relation between angular velocities~\eqref{eq:propeller_inputs} and control inputs~\eqref{eq:control_inputs} is given by:
\begin{equation}
\begin{bmatrix}
    T \\[3pt]
    \tau_x \\[3pt]
    \tau_y \\[3pt]
    \tau_z
\end{bmatrix}
=
\begin{bmatrix}
    k_F & k_F & k_F & k_F \\[3pt]
    -k_F L & -k_F L & k_F L & k_F L \\[3pt]
    -k_F L & k_F L & k_F L & -k_F L \\[3pt]
    -k_M & k_M & -k_M & k_M
\end{bmatrix}
\begin{bmatrix}
    \Omega_1^2 \\[3pt]
    \Omega_2^2 \\[3pt]
    \Omega_3^2 \\[3pt]
    \Omega_4^2
\end{bmatrix}.
\label{eq:thrust_torque_mapping}
\end{equation}
In~\eqref{eq:thrust_torque_mapping}, $k_F$ and $k_M$ are the thrust and moment coefficients of the propellers, while $L$ is the effective arm length, defined as the orthogonal distance from the nano-drone’s center of mass to each rotor projection on the body axes (see Figure~\ref{fig:crazyflie-convention}).
The first row expresses the total thrust as the sum of the four rotor contributions, while the remaining rows describe the roll, pitch, and yaw torques resulting from differential thrust and propeller drag.

\begin{remark}[Quadratic Rotor Model]\label{rem:quadratic_motor_model}
Equation~\eqref{eq:thrust_torque_mapping} relies on the assumption that each propeller produces thrust and torque proportional to the square of its angular speed, with proportionality coefficients $k_F$ and $k_M$, respectively.
Although this approximation suffices to illustrate the main dynamics and remains the standard in quadrotor modeling and control~\citep{mahony2012multirotor}, recent advances have produced more accurate descriptions of the aerodynamics~\citep{bauersfeld2021neurobem}.
These newer approaches, whether physics-based or data-driven, may offer a more accurate description of drone behavior, particularly during aggressive maneuvers.
\end{remark}

\section{Experimental Setup} \label{sec:setup}

\begin{figure}
    \centering
    \includegraphics[width=\linewidth]{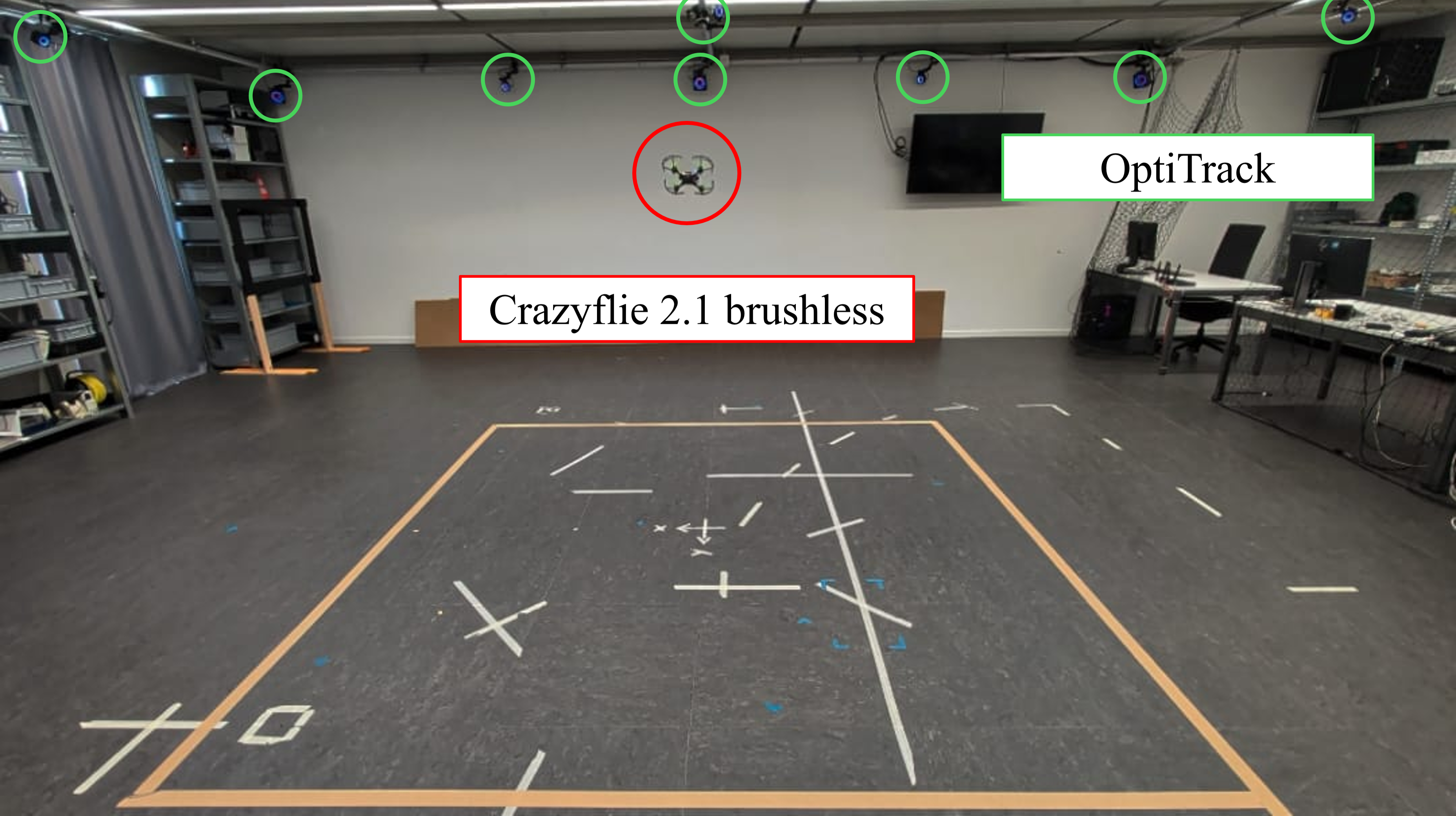}
    \caption{A Crazyflie 2.1 brushless flying in the motion-capture-equipped laboratory.}
    \label{fig:arena}
\end{figure}

All flights were carried out in the motion-capture arena described below and shown in  Figure~\ref{fig:arena}, with safety nets surrounding the test volume in an enclosed room to prevent external air flow during experiments.

\subsection{Crazyflie 2.1 Brushless Platform}

All experiments were conducted on a Crazyflie 2.1 Brushless nano–quadrotor equipped with a Flow-deck v2 and AI-deck expansion boards.
The platform provides on‑board computation, sensing, and communication through an STM32F405 flight MCU running the official Crazyflie firmware\footnote{\url{https://github.com/bitcraze/crazyflie-firmware}}, a Bosch BMI088 6-DOF Inertial Measurement Unit (IMU), and an NRF51822 \SI{2.4}{\giga\hertz} low-latency radio.
The vehicle is powered by four \SI{8}{\milli\meter} 10000KV brushless motors, driven by EFM8BB21 Electronic Speed Controllers (ESCs) with the Bluejay\footnote{\url{https://github.com/bitcraze/bluejay}} open-source ESC firmware.

The Flow-deck v2 also provides an ST VL53L1X Time-of-Flight laser-based altitude sensor and a PMW3901 optical flow sensor to measure horizontal displacement relative to the ground.
The AI-deck provides a Greenwaves Technologies GAP8 System-on-Chip for computation-intensive workloads~\citep{cereda2024trainonthefly} and an ublox NINA-W102 \SI{2.4}{\giga\hertz} Wi-Fi module for high-bandwidth communication.
A 1S \SI{350}{\milli\ampere\hour} LiPo battery powers the platform at a nominal voltage of \SI{3.7}{\volt}.
The total mass of the platform, including the battery, amounts to $m = \SI{45}{\gram}$. 
The inertia matrix $\mathbf{J} = \mathrm{diag}(J_x, J_y, J_z)$ is diagonal due to the symmetric construction. 
The numerical values of the diagonal entries, taken from the \emph{Crazyflow}\footnote{\url{https://github.com/utiasDSL/crazyflow}} simulator, are: $J_x = \num{2.3951e-5}, J_y = \num{2.3951e-5}, J_z = \num{3.2347e-5}$ \si{\kilo \gram \times \metre \squared}.

\subsection{Control System}

The quadrotor is controlled using a modified version of the geometric controller originally proposed by~\cite{mellinger2011minimum}, 
with integral control action added, as implemented in the Crazyflie’s official firmware and running at \SI{500}{\hertz} on the embedded STM32 MCU. 

Control setpoints are streamed from a laptop computer over the NRF51 radio using the Crazyswarm2\footnote{\url{https://imrclab.github.io/crazyswarm2/}} ROS package.
The controller tracks a desired position and yaw trajectory $\mathbf{p}_d(t), \psi_d(t)$ by computing the required total thrust $T$ and body torques $\bm{\tau} = [\tau_x, \tau_y, \tau_z]^\top$ that drive the vehicle toward the reference.
The outer position loop computes the desired force vector $$\mathbf{F}_{d} = \mathbf{K}_p(\mathbf{p}_d - \mathbf{p}) + \mathbf{K}_i\int (\mathbf{p}_d -\mathbf{p})\; dt + \mathbf{K}_d(\dot{\mathbf{p}}_d - \mathbf{v}) + m(\ddot{\mathbf{p}}_d - \mathbf{g}),$$ where $\mathbf{K}_p$, $\mathbf{K}_i$, and $\mathbf{K}_d$ are proportional, integral, and derivative diagonal matrix gains, while $\ddot{\mathbf{p}}_d$ is the desired feedforward acceleration.
The commanded thrust $T$ is then set as the component of $\mathbf{F}_{d}$ along the body's current $z$-axis: $T = \mathbf{F}_d \cdot z_{\body}$. 

The desired orientation $\mathbf{R}_d$ is then constructed with $z$-axis pointing in the direction of $\mathbf{F}_{d}$, while maintaining the reference yaw angle $\psi_d$.
The inner (attitude) loop runs directly on the manifold $\mathrm{SO}(3)$ and computes the angular velocity and torque commands required to track the desired attitude.
The attitude error is defined as $\mathbf{e}_R = \frac{1}{2}\left(\mathbf{R}_d^\top \mathbf{R}(q) - \mathbf{R}^\top(q) \mathbf{R}_d\right)^\vee$, where $(\cdot)^\vee$ is the standard \emph{vee operator} that unpacks the $3$ independent elements of skew-symmetric $3\times 3$ matrices into a vector with 3 components.
The angular velocity error is $\mathbf{e}_\omega = \bm{\omega} - \bm{\omega}_d$.
The control torque is then $$\bm{\tau} = -\mathbf{K}_R \mathbf{e}_R - \mathbf{K}_{R,i}\! \int\! \mathbf{e}_R\; dt - \mathbf{K}_\omega \mathbf{e}_\omega,$$ with $\mathbf{K}_R$, $\mathbf{K}_{R,i}$, and $\mathbf{K}_{\omega}$ the diagonal attitude, integral attitude, and angular-rate gain matrices.

\begin{table*}[h]
    \centering
    \caption{Dataset structure. \textbf{Dataset columns} denotes the name of the column in the \emph{csv} file of the provided benchmark dataset.}
    \resizebox{\linewidth}{!}{
    \begin{tabular}{cccccl}
    \toprule
    \textbf{Variable} & \textbf{Symbol} & \textbf{Unit} & \textbf{Ref. frame} & \textbf{Dataset columns} & \textbf{Measurement source} \\
    \midrule
    \vspace{0.4em} 
    Timestamp & $t$ & s & -- & \texttt{t} & STM32 hardware timer \\
    \multicolumn{4}{l}{\footnotesize\textbf{Inputs}} \\
    \midrule
    \vspace{0.4em} 
    Propeller speeds & $\mathbf\Omega$ & \si{\rad/\second} & -- &
    \texttt{m\{1,2,3,4\}\_rads} &
    Sensorless back-EMF zero crossing \\
    \multicolumn{4}{l}{\footnotesize\textbf{Outputs}} \\
    \midrule
    Position & $\mathbf{p}$ & \si{\meter} & world &
    \texttt{x, y, z} &
    Motion capture + Laser altimeter ($\texttt{z}$ only) \\
    Velocity & $\mathbf{v}$ & \si{\meter/\second} & world &
    \texttt{vx, vy, vz} &
    Optical flow ($\texttt{vx}$, $\texttt{vy}$ only) \\
    Orientation & $\mathbf{q}$ & quaternion & world &
    \texttt{qx, qy, qz, qw} &
    Motion capture \\
    Angular velocity & $\bm{\omega}$ & \si{\rad/\second} & body &
    \texttt{wx, wy, wz} &
    Gyroscope \\
    \addlinespace[0.6em]
    \multicolumn{4}{l}{\footnotesize\textbf{Additional}} \\
    \midrule
    \vspace{0.4em} 
    Specific acceleration & $\mathbf{a}^{\mathrm{IMU}}$ & \si{\meter/\second^2} & body &
    \texttt{a\{x,y,x\}\_body} &
    Accelerometer. It measures $(\mathbf{a} - \mathbf{g})_\body$ \\
    \bottomrule
    \end{tabular}
    }
    \label{tab:dataset_structure}
\end{table*}

\subsection{Sensing and State Estimation}\label{sec:state_estimation}

Data were recorded using a combination of onboard sensors and an external motion-capture system. 
Ground-truth position and orientation were obtained from an \emph{OptiTrack} setup with 18 cameras covering a 6$\times$6$\times$\SI{2.5}{\meter} volume.

The onboard sensing suite includes: ($i$) a 3-axis gyroscope and accelerometer for angular velocity and specific acceleration, ($ii$) a PMW3901 optical flow module for horizontal velocity estimation, ($iii$) a VL53L1x time-of-flight sensor for altitude, and ($iv$) propeller angular velocity measurements obtained from the ESCs via back-EMF zero-crossing detection.
These angular velocities values are transmitted to the STM32 flight controller using the bidirectional DSHOT protocol.

State estimation is performed online by an Extended Kalman Filter (EKF)~\citep{mueller2015fusing, mueller2017covariance}, executed at \SI{100}{\hertz} in the Crazyflie firmware.  
The filter fuses IMU, optical flow, time-of-flight, and motion-capture measurements to provide a full 6-DoF pose estimate, used as feedback by the low-level controller.

Because the official Crazyflie firmware supports only unidirectional DSHOT, we developed a bidirectional driver to enable motor-speed telemetry. 
The implementation requires precise synchronization of transmit and receive phases across the four motors. 
We open-source this driver and are integrating it into the official firmware.
\section{Dataset and Evaluation Protocol} \label{sec:experiments}

This section describes the trajectories used to collect the dataset.
It provides details on the data itself, as well as the evaluation settings adopted to assess model identification performance on the proposed benchmark.
Table~\ref{tab:dataset_structure} summarizes the structure of the recorded dataset, which includes motor speeds $\mathbf{u} = [\Omega_1\ \Omega_2\ \Omega_3\ \Omega_4]^\top$, position and orientation (in quaternion form), angular velocity, and linear acceleration.

\subsection{Data Preprocessing} \label{sec:preprocessing}

All data is streamed in real-time using high-throughput Wi-Fi streaming through our internally developed NanoCockpit framework and recorded in a \texttt{rosbag} file. 
This recording is asynchronous and timestamped at the source using the onboard \textsc{STM32} hardware clock.
Since the motion-capture system and the flight controller operate at different frequencies and with independent clocks, each experiment required careful post-processing to obtain a time-aligned, uniformly sampled dataset.

\paragraph{Timestamp synchronization}
Raw data from the Crazyflie firmware and the motion-capture system are first aligned in time by estimating their relative delay.
We compute the lag that maximizes the cross-correlation between the position signals $(x,y,z)$ from the OptiTrack and the corresponding onboard estimates, and subtract the average of the three estimated lags to reduce noise sensitivity.
After alignment, all channels are resampled to a common  $100\,\mathrm{Hz}$ rate using interpolation.
Because the embedded hardware clock may exhibit irregular sampling intervals, we additionally perform a \emph{retiming}  with nearest-neighbor interpolation, to guarantee that all data after processing  are uniformly sampled. This also compensates for any potentially missing packets by assigning them the closest available measurement.

\paragraph{Flight-segment extraction}
Each experiment begins with an initial stationary phase, during which the trajectory is uploaded, and the vehicle takes off.
We automatically extract the flight segment by selecting the time interval in which the reference position is non-zero. This produces consistent flight windows across repetitions without manual trimming.

\paragraph{Motor–acceleration alignment}
Motor speed telemetry is affected by an additional delay in communication and actuation.
To compensate for this, we align the motor commands to the inertial measurements by maximizing the cross-correlation between the vertical body acceleration $a_z$ and the instantaneous total thrust, approximated as the sum of squared motor speeds.
This yields a robust estimate of the command–actuation delay, ensuring consistent temporal alignment between the thrust input and measured accelerations.

\paragraph{Filtering pipeline}
To reduce measurement noise while preserving dynamic content relevant for system identification, all continuous-valued signals are filtered using zero-phase fourth-order Butterworth low-pass filters.
Since different physical quantities exhibit different spectral characteristics, we assign a dedicated cutoff frequency to each signal group.
Short gaps in the raw data (up to 10 samples) are filled via bounded interpolation, while longer gaps remain untouched. 
Quaternion signals are 
mapped to the rotation-vector representation via the logarithmic map, low-pass filtered component-wise, and mapped back using the exponential map.  
This procedure ensures that filtering operates in a linear space, avoids quaternion drift or non-unit norms, and preserves the manifold structure of $\mathrm{SO}(3)$, which would be violated by direct filtering in quaternion coordinates.
Table~\ref{tab:filter_cutoffs} summarizes the cutoff frequencies applied to each variable group.

\begin{table}[t]
    \centering
    \caption{Low-pass filter cutoff frequencies for different signal groups ($f_s = 100$\,Hz, Butterworth order = 4).}
    \label{tab:filter_cutoffs}
    \renewcommand{\arraystretch}{1.2}
    \resizebox{\linewidth}{!}{
    \begin{tabular}{l c}
        \toprule
        \textbf{Signal group} & \textbf{Cutoff frequency [Hz]} \\
        \midrule
        Position $(x,y,z)$ & 10 \\
        Linear and angular velocities $(v,\omega)$ & 18 \\
        Linear accelerations $(a_x,a_y,a_z)$ & 25 \\
        Motor speeds & 20 \\
        Quaternions (via log-map) & 12 \\
        \bottomrule
    \end{tabular}
    }
\end{table}

\subsection{Trajectory Design}

\begin{figure*}[t]
    \centering
    \begin{subfigure}[t]{0.24\textwidth}
        \centering
        \subcaption{\href{https://github.com/idsia-robotics/nanodrone-sysid-benchmark/blob/main/animations/square_run1.gif}{Square}}
        \includegraphics[width=\linewidth]{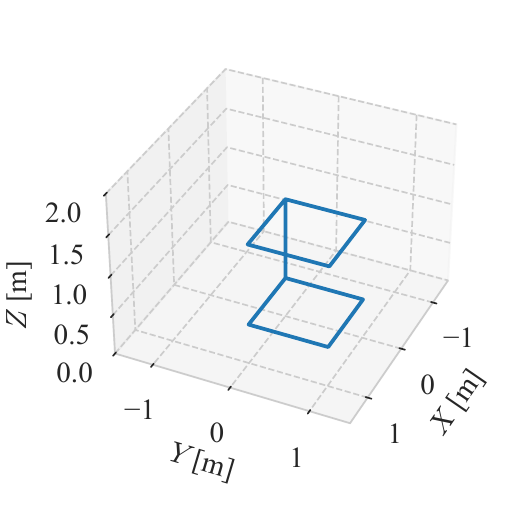}
        \label{fig:traj_square_ideal}
    \end{subfigure}\hfill
    \begin{subfigure}[t]{0.24\textwidth}
        \centering
        \subcaption{\href{https://github.com/idsia-robotics/nanodrone-sysid-benchmark/blob/main/animations/random_run1.gif}{Random}}
        \includegraphics[width=\linewidth]{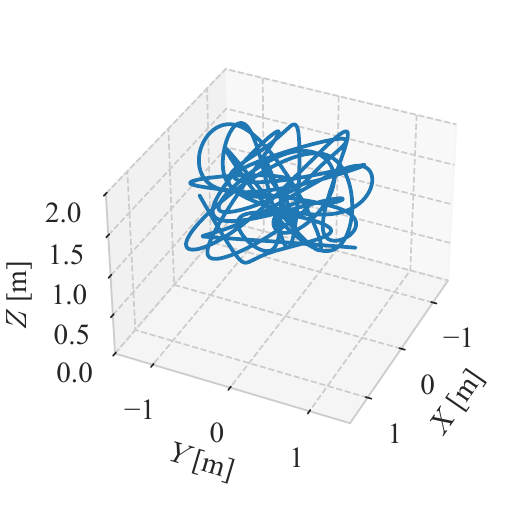}
        \label{fig:traj_random_ideal}
    \end{subfigure}\hfill
    \begin{subfigure}[t]{0.24\textwidth}
        \centering
        \subcaption{\href{https://github.com/idsia-robotics/nanodrone-sysid-benchmark/blob/main/animations/melon_run1.gif}{Melon}}
        \includegraphics[width=\linewidth]{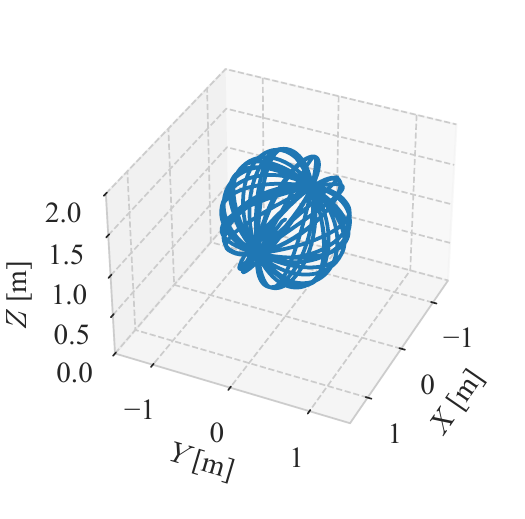}
        \label{fig:traj_melon_ideal}
    \end{subfigure}\hfill
    \begin{subfigure}[t]{0.24\textwidth}
        \centering
        \subcaption{\href{https://github.com/idsia-robotics/nanodrone-sysid-benchmark/blob/main/animations/chirp_run1.gif}{Chirp}}
        \includegraphics[width=\linewidth]{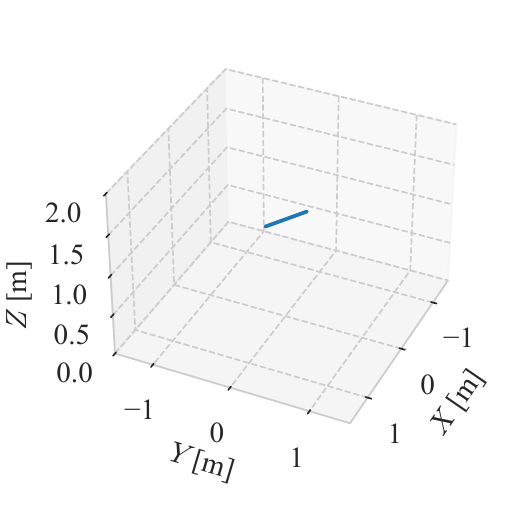}
        \label{fig:traj_chirp_ideal}
    \end{subfigure}

    \vspace{0.cm}

    \begin{subfigure}[t]{0.24\textwidth}
        \centering
        \includegraphics[width=\linewidth, trim=10 165 165 40, clip]{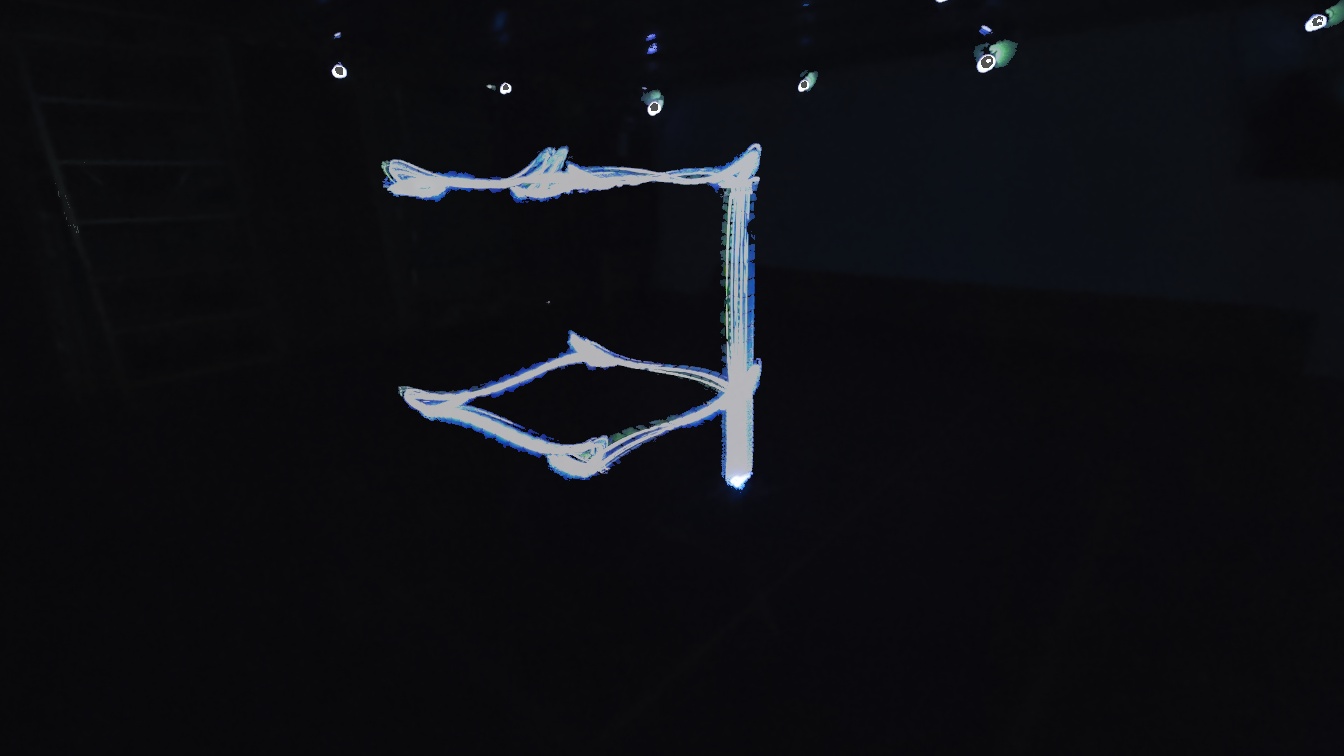}
        \label{fig:traj_square_real}
    \end{subfigure}\hfill
    \begin{subfigure}[t]{0.24\textwidth}
        \centering
        \includegraphics[width=\linewidth, trim=0 120 0 0, clip]{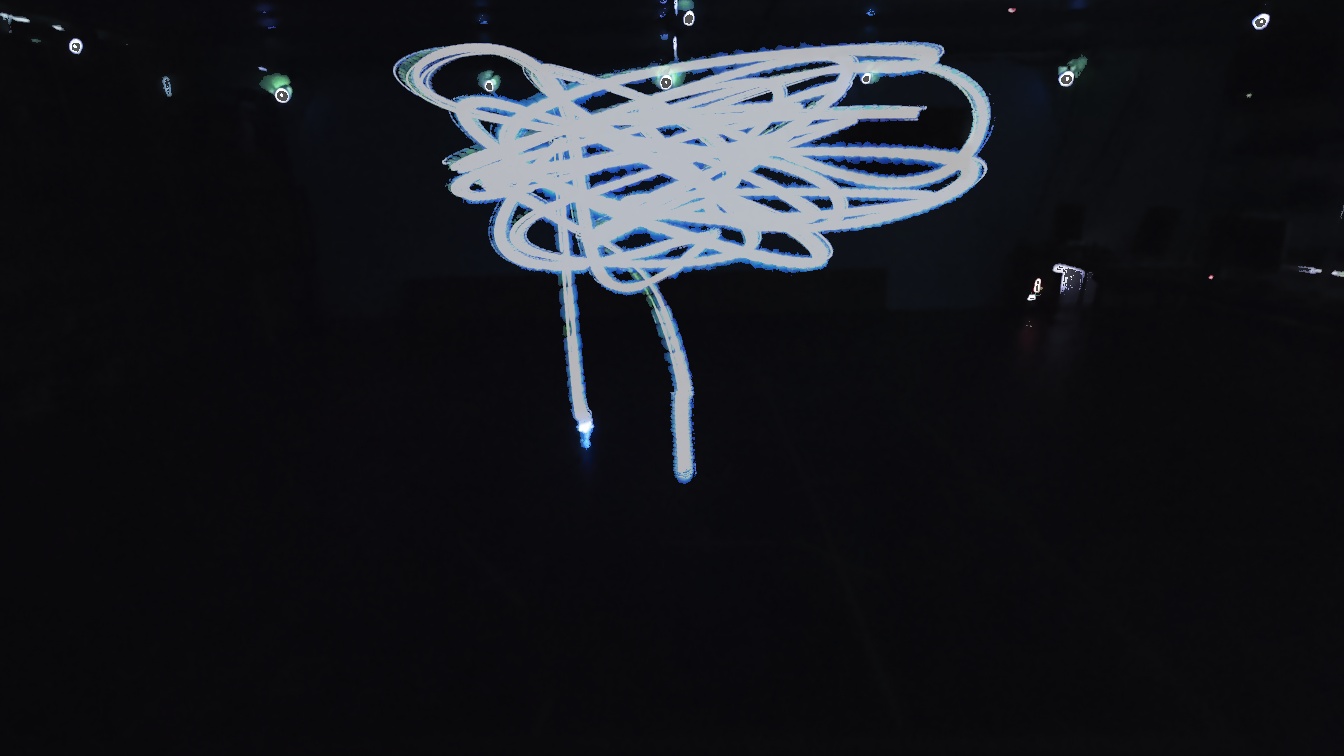}
        \label{fig:traj_random_real}
    \end{subfigure}\hfill
    \begin{subfigure}[t]{0.24\textwidth}
        \centering
        \includegraphics[width=\linewidth, trim=0 120 0 0, clip]{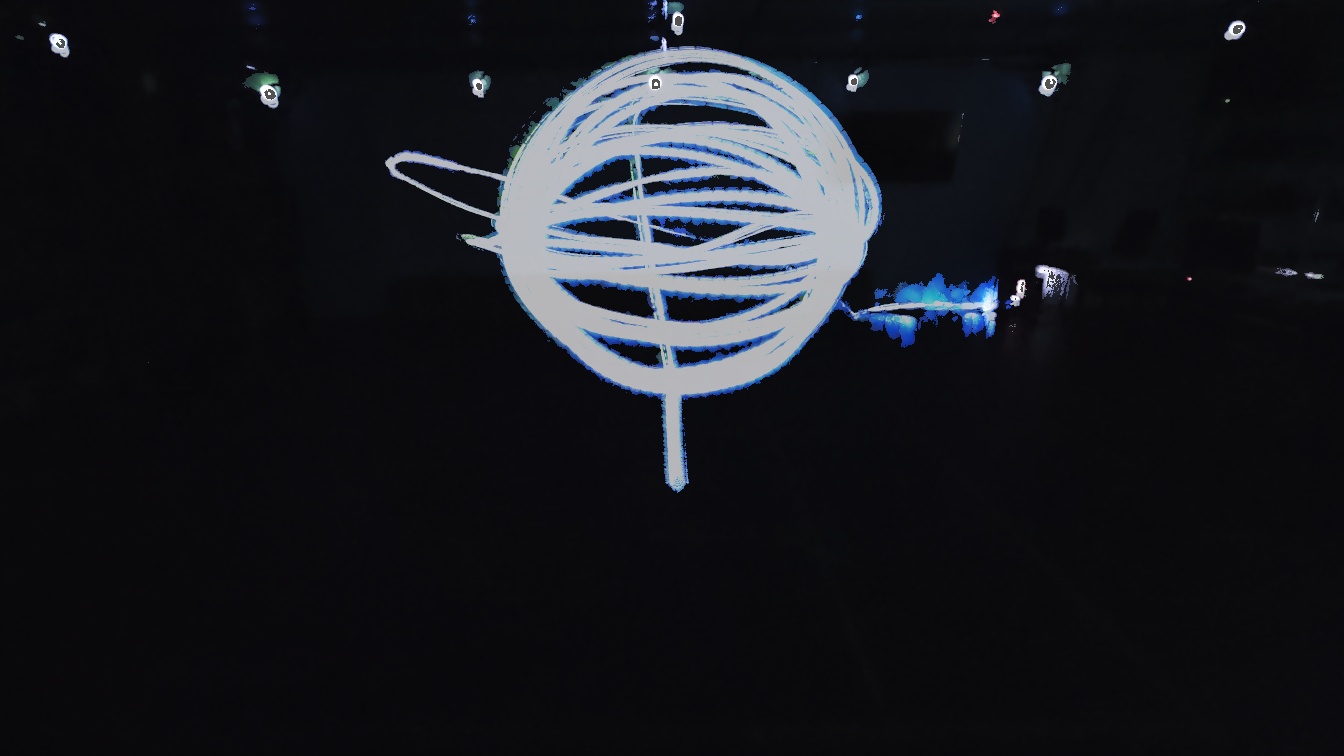}
        \label{fig:traj_melon_real}
    \end{subfigure}\hfill
    \begin{subfigure}[t]{0.24\textwidth}
        \centering
        \includegraphics[width=\linewidth, trim=0 120 0 0, clip]{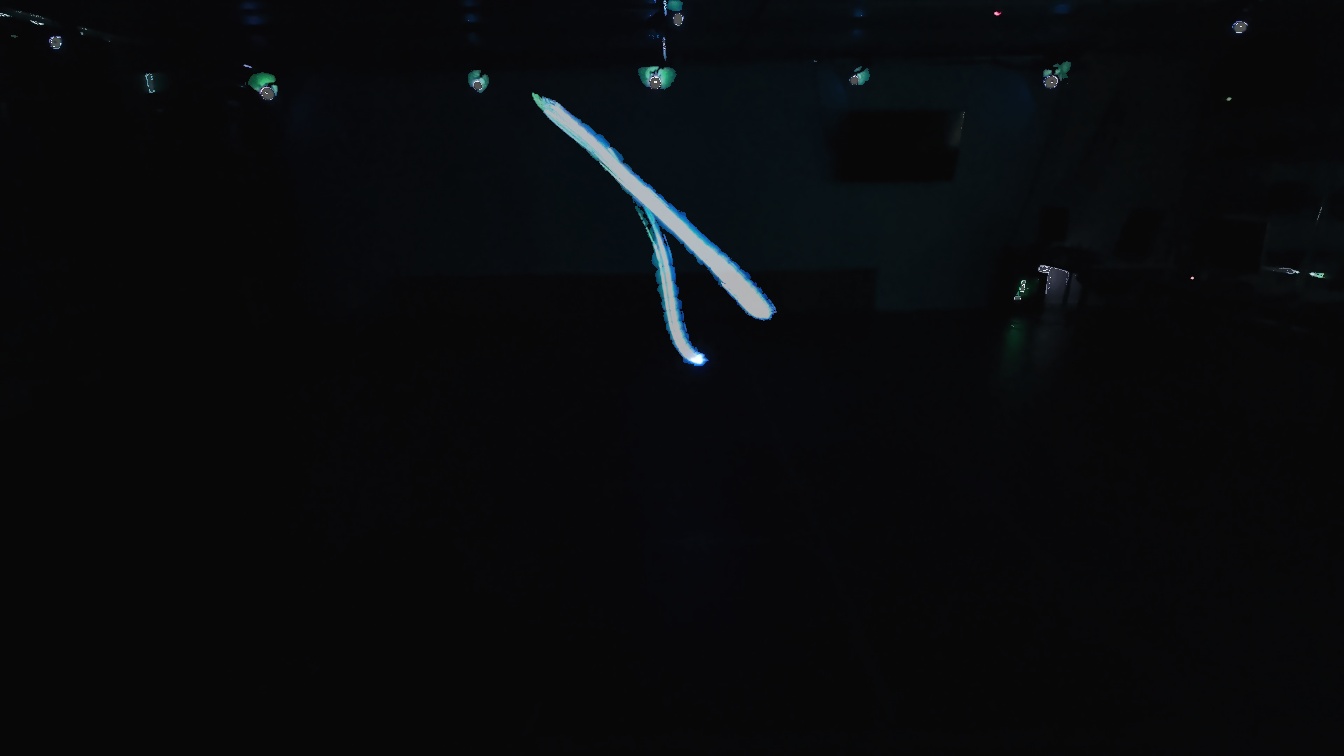}
        \label{fig:traj_chirp_real}
    \end{subfigure}

    \caption{
        \textbf{Reference trajectories and corresponding long-exposure flight traces.} Top row: ideal 3D reference trajectories generated in Python. Bottom row: long-exposure images of Crazyflie~2.1 executing the same trajectories in the motion-capture arena. (Click sub-captions in the PDF viewer to see the animations).
    }
    \label{fig:combined_trajectories}
\end{figure*}

Four reference trajectories (named \emph{Square}, \emph{Random}, \emph{Melon}, and \emph{Chirp}) were designed to excite the quadrotor dynamics over a broad range of operating conditions while remaining bounded and repeatable. 
All trajectories were sampled at \SI{100}{\hertz} and executed at least three times (referred to as \textit{runs}); Square, Random, and Chirp include a 4th run, for a total of 15 flights. 
The reference paths are illustrated in Figure~\ref{fig:combined_trajectories}, and their main statistics are summarized in Table~\ref{tab:reference_trajectories}.
Together, these trajectories provide a balanced set of structured, random, and frequency-rich motions for identification and validation.

\paragraph{Square}
A structured path consisting of two horizontal squares of \SI{1}{\meter} side length, stacked 1~m apart in height. 
Each segment is executed in \SI{1}{\second}, followed by \SI{1}{\second} of hovering to allow stabilization between transitions. 
The motion along each edge is generated using \emph{minimum-snap interpolation}, ensuring continuity in position, velocity, and acceleration. 
After completing the lower loop, the vehicle ascends to the upper plane, repeats the same pattern, and finally returns to the starting height. 
This trajectory primarily excites roll and pitch dynamics independently, providing a repeatable planar motion for controller validation.

\paragraph{Random-points}
A \SI{60}{\second} trajectory composed of 52 way-points uniformly sampled within a box of size \(1{\times}1{\times}0.5~\mathrm{m}\) centered at \((0,0,1.5)\).
Positions are interpolated using cubic splines to generate smooth velocity and acceleration profiles. 
The resulting motion is irregular and non-periodic, with broadband excitation arising from the random spacing of consecutive points.

\paragraph{Melon}
A smooth, periodic 65~s motion combining elliptical oscillations in the local $x$--$z$ plane with a slow rotation of the plane about the \(x\)-axis $\omega_{\text{plane}} = \SI{0.4}{\rad / \second}$. 
The ellipse radii are fixed to \(\SI{0.75}{\metre}\), and the angular velocity to \(\omega_{\text{circle}} = \SI{2.5}{\rad / \second}\). 
After the main segment, the drone returns smoothly to the center using a minimum-jerk transition. 
This trajectory generates coupled translational and rotational motion, providing significant excitation of the attitude dynamics.

\paragraph{Chirp}
A \SI{60}{\second} trajectory in which the three position axes are simultaneously driven by independent sinusoidal signals of amplitude \(0.5~\mathrm{m}\). 
The instantaneous frequencies are linearly increased from \SI{0.1}{\hertz} to \SI{0.5}{\hertz}, resulting in a rich spectral content across multiple frequency bands. 
This trajectory yields persistent excitation of both translational and rotational modes.

\paragraph{Train-test split}
Experimental data collected from the \textit{Square}, \textit{Random}, and \textit{Chirp} trajectories are used for training, while the \textit{Melon} trajectory is reserved exclusively for testing,
yielding \num{55999} training samples (74.2\%) and \num{19497} testing samples (25.8\%).
We decided to exclude the measured acceleration from the output variables, as it is not a state of the system.

\begin{remark}[Zero yaw rate]\label{rem:yaw_rate}
All trajectories were generated with a fixed yaw orientation, i.e., the reference yaw satisfies $\psi^{\mathrm{ref}} = 0$ throughout the experiments. 
This choice simplifies the control problem by decoupling yaw from roll and pitch dynamics. 
It reflects common practice in the identification of drone dynamics, where the focus is typically on modeling translational and attitude dynamics under a constant heading \citep{bauersfeld2021neurobem}.
\end{remark}

\begin{table}[h!]
\centering
\caption{Benchmark statistics for real quadrotor trajectories. Reported values are averaged across runs.}
\begin{tabular}{l|ccccc}
\toprule
&
\rotatebox{90}{Duration [s]} &
\rotatebox{90}{$v_{\mathrm{mean}}$ [m/s]} &
\rotatebox{90}{$v_{\mathrm{max}}$ [m/s]} &
\rotatebox{90}{$a_{\mathrm{max}}$ [m/s²]} &
\rotatebox{90}{Max tilt [°]} \\
\midrule
Square & 19.00 & 0.83 & 2.60 & 12.53 & 42.95 \\
Random & 60.00 & 1.11 & 2.69 & 8.30 & 43.23 \\
Melon & 65.00 & 1.44 & 3.06 & 10.91 & 56.07 \\
Chirp & 60.00 & 0.88 & 3.02 & 10.13 & 62.69 \\
\bottomrule
\end{tabular}
\label{tab:reference_trajectories}
\end{table}

\subsection{Evaluation and Metrics}
Let $\mathbf{y}_t$ denote the set of the \textbf{Outputs} variables in Table~\ref{tab:dataset_structure} at time~$t$:
\begin{equation}
    \label{eq:y_def}
    \mathbf{y}_t
    =
    \begin{bmatrix}
        \mathbf{p}_t^{\top} \
        \mathbf{v}_t^{\top} \
        \mathbf{q}_t^{\top} \
        \bm{\omega}_t^{\top}
    \end{bmatrix}^{\top}
    \in \mathbb{R}^{13},
\end{equation}
where $\mathbf{p}_t \in \mathbb{R}^3$ is the position, $\mathbf{v}_t \in \mathbb{R}^3$ the linear velocity,
$\mathbf{q}_t \in \mathbb{R}^4$ the unit quaternion representing attitude, and $\bm{\omega}_t \in \mathbb{R}^3$ the angular velocity. 
The benchmark evaluates how accurately the model predicts the evolution of the output variables over rolling windows of maximum length $H=50$ (corresponding to \SI{0.5}{\second} at \SI{100}{\hertz}).
At test time, the model receives the true variables $\mathbf{y}_t$ at the beginning of the window
and the full input sequence $\{\mathbf{u}_{t+k}\}_{k=0}^{H-1}$, and must predict the future
states $\{\hat{\mathbf{y}}_{t+k}\}_{k=1}^{H}$ without additional state corrections.

Metrics are computed on the held-out \textit{Melon} trajectory $\{(\mathbf{y}_t, \mathbf{u}_t)\}_{t=0}^{T_{\mathrm{test}}-1}$.
For each admissible start time $t \in \mathcal{T} := \{0, 1, \dots, T_{\mathrm{test}} - H -1 \},$\ the model produces a multi-step open-loop prediction:
\begin{equation*}
    \hat{\mathbf{y}}_{t+1:t+H}
    = 
    M\!\left(\mathbf{y}_t, \mathbf{u}_{t:t+H-1}\right).
\end{equation*}
Thus, each start time \(t \in \mathcal{T} \) yields a set of \(H\) predictions \(\{\hat{\mathbf{y}}_{t+1|t},\dots,\hat{\mathbf{y}}_{t+H|t}\}\), which is compared to the corresponding ground-truth sequence $\mathbf{y}_{t+1:t+H}$ using error metrics defined below.

For positions, linear velocities, and angular velocities, we define the instantaneous Euclidean prediction error at horizon~$h$ as:
\begin{alignat}{2}
    e_{p,t,h} &= \bigl\|\,\mathbf{p}_{t+h} - \hat{\mathbf{p}}_{t+h\mid t}\,&&\bigr\|_2,\\[2mm]
    e_{v,t,h} &= \bigl\|\,\mathbf{v}_{t+h} - \hat{\mathbf{v}}_{t+h\mid t}\,&&\bigr\|_2,\\[2mm]
    e_{\omega,t,h} &= \bigl\|\,\bm{\omega}_{t+h} - \hat{\bm{\omega}}_{t+h\mid t}\,&&\bigr\|_2 .
\end{alignat}

For the orientation, we adopt as error metric the geodesic distance on $\mathrm{SO}(3)$ \citep{huynh2009metrics} between the true attitude $q_{t+h}$ and its prediction $\hat{q}_{t+h\mid t}$. The relative rotation is \[q_{\mathrm{rel},t,h} = q_{t+h}^{-1} \otimes \hat{q}_{t+h\mid t} = \begin{bmatrix} \mathbf{q}_{v} \\ q_{w} \end{bmatrix}, \] with vector part $\mathbf{q}_v$ and scalar part $q_w$.  
The corresponding axis--angle rotation error is given by:
\begin{equation}
    e_{R,t,h}
    =
    2\,\operatorname{atan2}\left( \bigl\| \mathbf{q}_{v} \bigr\|,\; q_{w} \right),
\end{equation}
which yields the geodesic orientation error in radians.

We evaluate the prediction accuracy using the mean absolute error per-prediction horizon, defined as:
\begin{equation}
    \mathrm{MAE}_{(\cdot),h}
    = \frac{1}{|\mathcal{T}|\,}
      \sum_{t\in\mathcal{T}}
      e_{(\cdot),t,h},
      \quad h = 1,\dots,H,
\end{equation}
where $(\cdot)\in\{p,\,v,\,\omega,\,R\}$ indicates quantity of interest. 
For a compact comparison across models, we also report an aggregated simulation error metric, defined as the average prediction error accumulated over a $H=50$-step open-loop rollout:
\begin{equation}
\mathrm{MAE}_{(\cdot),1:H} = \sum_{h=1}^{H}
\mathrm{MAE}_{(\cdot),h}.
\end{equation}
This 50-step simulation error summarizes the model’s open-loop performance over a finite horizon, providing a compact indicator of overall simulation quality.
For reproducibility, the accompanying repository provides ready-to-use implementations of all the evaluation metrics defined above.
\section{Results} \label{sec:results}

All models were implemented in \texttt{PyTorch} with differentiable integration and trained using the data splits described in Section~\ref{sec:preprocessing}. 
Preprocessing, model training, and evaluation were performed on a laptop equipped with an Intel\textsuperscript{\textregistered} Core\texttrademark{} i9--14900HX CPU (24~cores, 32~threads), \SI{32}{\giga\byte} of RAM, and an NVIDIA GeForce RTX~4070 Laptop GPU, running Windows~11 (build~22621). 
The complete implementation is publicly available at 
\url{https://github.com/idsia-robotics/nanodrone-sysid-benchmark}. 

\subsection{Data representation for black-box model learning} \label{sec:data_representation}

As introduced in Section~\ref{sec:settings}, the quadrotor is actuated by four brushless motors whose angular velocities form the input vector $\mathbf{u} \in \mathbb{R}^{4}$, see~\eqref{eq:propeller_inputs}. Selecting a subset of relevant variables from the measurement vector $\mathbf{y}$ and expressing them in a representation tailored for learning can simplify the regression problem and improve numerical
stability.

For example, a direct regression on quaternion orientations $\mathbf{q} = [q_x, q_y, q_z, q_w]^\top$ is undesirable due to the unit-norm constraint and the double-covering property of quaternions, both of which introduce discontinuities in the learning problem.
To avoid these issues, we map each quaternion into its equivalent \emph{rotation-vector} (axis--angle) form
using the logarithmic map on $\mathrm{SO}(3)$:
\begin{equation}
    \label{eq:quat_to_so3}
    \mathbf{r}
    = 2\,\operatorname{atan2}\!\bigl(\|\mathbf{q}_v\|,\; q_w\bigr)\,
      \frac{\mathbf{q}_v}{\|\mathbf{q}_v\|},
\end{equation}
where $\mathbf{q}_v = [q_x, q_y, q_z]^\top$ is the vector part of the quaternion.
The resulting vector $\mathbf{r} \in \mathbb{R}^3$ provides a minimal and topologically consistent parameterization of orientation, free of normalization constraints and sign ambiguities~\citep{sola2021microlie}.

Using this representation, the output vector used for learning is defined as:
\begin{equation}
    \mathbf{\tilde{y}} =
    \begin{bmatrix}
        \mathbf{p}^\top \; \mathbf{v}^\top \; \mathbf{r}^\top \; \bm{\omega}^\top
    \end{bmatrix}^\top
    \in \mathbb{R}^{12},
\end{equation}
where $\mathbf{p} = [p_x, p_y, p_z]^\top$ is the position in the world frame $\mathcal{W}$, $\mathbf{r}$ is the rotation vector obtained from the quaternion orientation, $\mathbf{v} = [v_x, v_y, v_z]^\top$ is the linear velocity in $\mathcal{W}$, and $\bm{\omega} = [\omega_x, \omega_y, \omega_z]^\top$ is the body-frame angular velocity.

Training is performed using fixed-length input--output windows of horizon \(H = 50\) samples (0.5~s).
For each trajectory \(j \in \mathcal{J}_{\mathrm{train}} = \{\text{Square}, \text{Random}, \text{Multisine}\}\), with \(T_j\) denoting the total number of time samples, the dataset is constructed as:
\begin{equation}
\mathcal{D}_{\mathrm{train}}
= \bigcup_{j \in \mathcal{J}_{\mathrm{train}}}
\left\{
\big(\mathbf{X}_t, \mathbf{Y}_t\big)
\right\}_{t=0}^{T_j - H},
\end{equation}
where each pair corresponds to a temporal window:
\begin{align}\label{eq:dataset_representation}
\mathbf{X}_t &= \big\{ \mathbf{\tilde{y}}_t,\, \mathbf{u}_{t:t+H-1} \big\}, \\
\mathbf{Y}_t &= \big\{ \mathbf{\tilde{y}}_{t+1:t+H} \big\}.
\end{align}
The same construction is applied to the \textit{Melon} trajectory to obtain the test set \(\mathcal{D}_{\mathrm{test}}\).

\subsection{Na\"ive Model}

As a reference for evaluating the predictive performance of the proposed models, we consider a constant predictor as a baseline. 
This model assumes that the system output remains constant over the prediction horizon:
\begin{equation}
    \hat{\mathbf{y}}_{t+h|t} = \mathbf{\tilde{y}}_t,
    \quad h = 1,\dots,H,
    \label{eq:naive_baseline}
\end{equation}
where \(\hat{\mathbf{y}}_{t+h|t}\) denotes the prediction of the output vector \(\mathbf{\tilde{y}}_{t+h|t}\) (used for model training)  \(h\) steps ahead given the current measurement \(\mathbf{\tilde{y}}_t\).
Although this approach neglects the system dynamics entirely, it establishes a meaningful reference level for short-horizon predictions, since it captures the strong temporal correlation often present in high-rate sampled data. 
Models that do not outperform this baseline for small horizons \(h\) can be considered ineffective, while a consistent improvement indicates that the model has successfully captured meaningful system dynamics.

\subsection{Physical model} \label{subsec:phys_model}

The nominal physical model is based on the continuous-time quadrotor dynamics given in~\eqref{eq:quad_dynamics}, together with the quadratic aerodynamics equations in \eqref{eq:thrust_torque_mapping}.
The system state is \(\mathbf{x} = \mathbf{y} =[\mathbf{p}^\top, \mathbf{v}^\top, \mathbf{q}^\top, \bm{\omega}^\top]^\top\), while the control input \(\mathbf{u}\) corresponds to the four propellers' rotational speeds.

To obtain a discrete-time model  suitable for multi-step prediction and simulation, the continuous-time dynamics are numerically integrated over each sampling interval \(\Delta t\) using a fourth-order Runge–Kutta (RK4) scheme.
Starting from the current measurement \(\mathbf{y}_t\), the physical model recursively predicts the future outputs as:
\begin{align}
    \hat{\mathbf{y}}_{t|t} &= \mathbf{\tilde{y}}_t, \\[3pt]
    \hat{\mathbf{y}}_{t+h|t} &=
    f_{\text{phys}}\!\big(
        \hat{\mathbf{y}}_{t+h-1|t},
        \mathbf{u}_{t+h-1}
    \big),
    \quad h = 1,\dots,H,
    \label{eq:phys_discrete}
\end{align}
where \(f_{\text{phys}}\) denotes the discrete-time transition function implementing the full rigid-body and motor dynamics. Notice that \(f_{\text{phys}}\) contains the transformation from axes-angle convention to quaternions and the inverse, to keep consistency with data representation in \eqref{eq:dataset_representation}.
This model requires the identification of the propeller thrust and drag coefficients, \(k_F\) and \(k_M\), which map the normalized motor commands to physical thrust and torque values.

\paragraph{Parameter Estimation}
The total thrust and torques generated by the propellers are modeled according to the quadratic aerodynamics model in~\eqref{eq:thrust_torque_mapping}, which depends on the unknown coefficients $k_F$ and $k_M$.
To estimate the thrust coefficient $k_F$, it is convenient to express the translation dynamics (second equation of~\eqref{eq:quad_dynamics}) in body coordinates: $ \mathbf{f}_\body = m (\mathbf{a} - \mathbf{g})_\body = [0 \; 0\; T]^\top.$
We replace $(\mathbf{a} - \mathbf{g})_\body$ with the accelerometer reading $\mathbf{a}^{\mathrm{IMU}}$, which directly measures \emph{specific} acceleration in body frame coordinates: $\mathbf{f}_\body \approx \tilde {\mathbf{f}}_\body = m \mathbf{a}^{\mathrm{IMU}}$.
By considering the force balance along the body's $z$-axis and by substituting the total thrust $T$ with its constitutive equation from the first row of~\eqref{eq:thrust_torque_mapping}, we get:
\begin{equation}
    \tilde f_{\body, z} = m a^{\mathrm{IMU}}_{z} = k_F(\Omega_1^2 + \Omega_2^2+\Omega_3^2+\Omega_4^2).
    \label{eq:thrust_identification}
\end{equation}
The equation above depends on measured quantities and the unknown parameter \(k_F\), allowing for the estimation of the latter.
Least-squares fittings over the training dataset yields $k_F = \SI{3.72e-8}{\newton \second^{2} \per \radian^{2}}$.

\begin{figure}[t]
    \centering
    \begin{subfigure}{\linewidth}
        \centering
        \includegraphics[width=\linewidth]{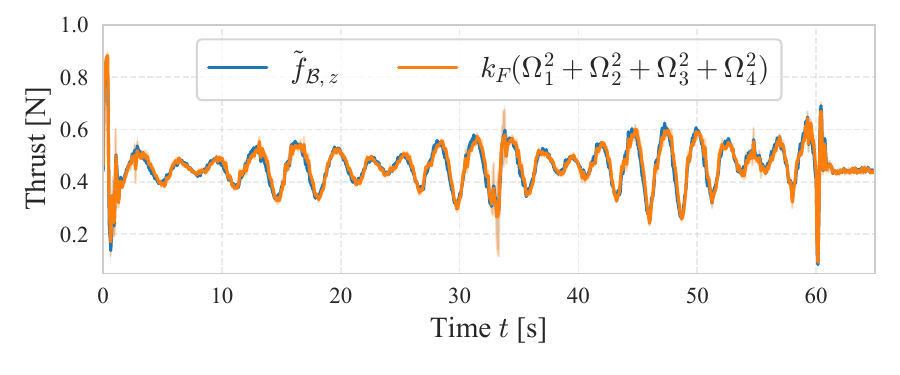}
        \label{fig:thrust_timeseries_mean_std}
    \end{subfigure}

    \vspace{-1cm}

    \begin{subfigure}{\linewidth}
        \centering
        \includegraphics[width=\linewidth]{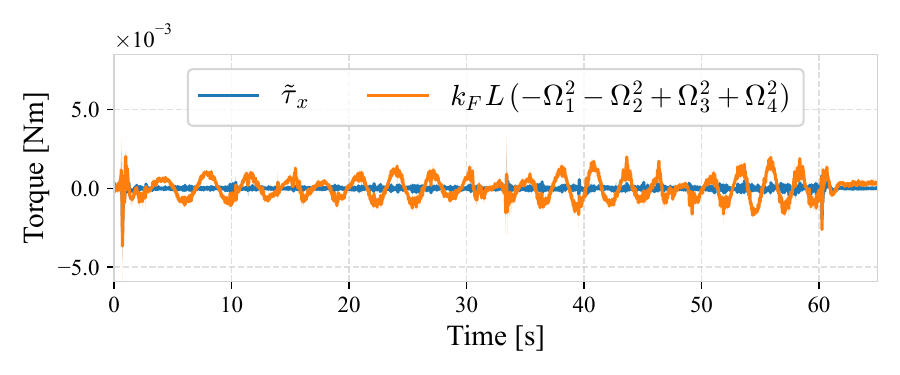}
        \label{fig:torque_x_timeseries_mean_std}
    \end{subfigure}

    \vspace{-1cm}

    \begin{subfigure}{\linewidth}
        \centering
        \includegraphics[width=\linewidth]{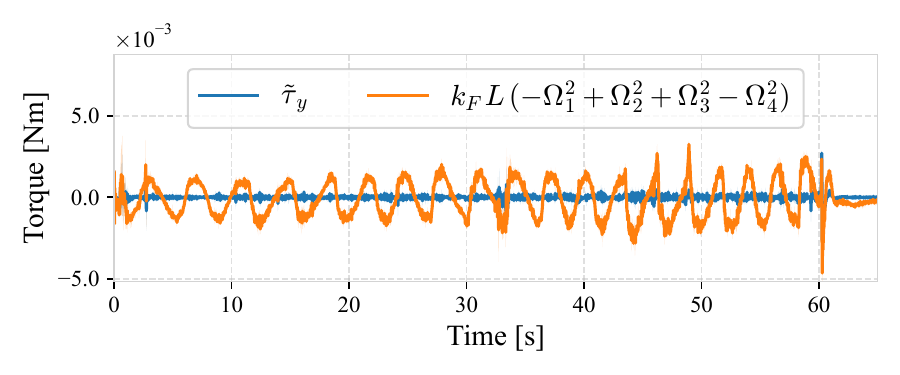}
        \label{fig:torque_y_timeseries_mean_std}
    \end{subfigure}

    \vspace{-1cm}

    \begin{subfigure}{\linewidth}
        \centering
        \includegraphics[width=\linewidth]{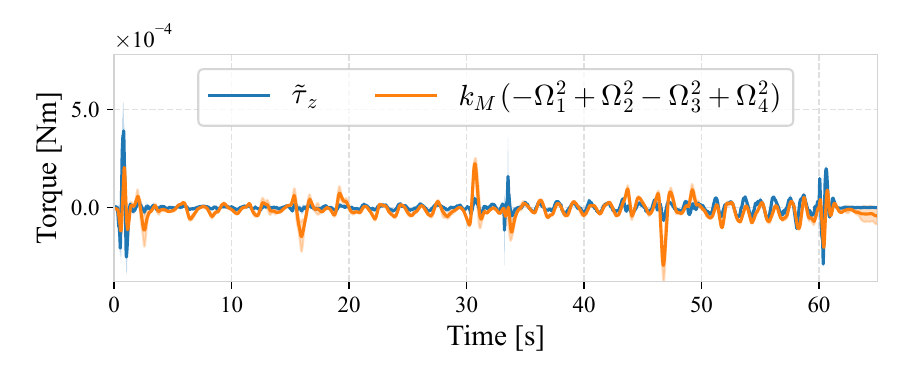}
        \label{fig:torque_z_timeseries_mean_std}
    \end{subfigure}

    \caption{Comparison between measured and model-predicted thrust and yaw torque
    on the \emph{Melon} test trajectory (run 1), averaged over three runs, using the
    identified aerodynamic coefficients $k_F$ and $k_M$.}
    \label{fig:thrust_yaw_comparison}
\end{figure}

A similar procedure is used to estimate the torque coefficient $k_M$. The ground-truth torque signal $\tilde {\bm{\tau}}$ is first reconstructed as $$\tilde {\bm{\tau}} = \mathbf{J} \dot{\bm{\omega}} + \bm{\omega} \times (\mathbf{J} \bm{\omega}),$$ following the rigid body angular dynamics in the last equation of $\eqref{eq:quad_dynamics}$. 
The reconstruction is based on the known inertia matrix $\mathbf{J}$, the angular velocity $\bm{\omega}$ measured by the gyroscope, and a finite difference approximation of the angular acceleration $\dot{\bm{\omega}}$ obtained from the same signal.

Considering the constitutive equation for $\tau_z$ at the last row of~\eqref{eq:thrust_torque_mapping}, we get:
\begin{equation}
    \tilde \tau_z = k_M\, (-\Omega_1^2 + \Omega_2^2 - \Omega_3^2 + \Omega_4).
    \label{eq:torque_balance}
\end{equation}
The parameter \(k_M\) is then estimated via least-squares regression over the training dataset based on~\eqref{eq:torque_balance}. 
We obtain $k_M = \SI{7.73 e-11}{\newton \metre \second^{2} \per \radian^{2}}$.

\begin{figure}[t]
    \centering
    \includegraphics[width=\linewidth]{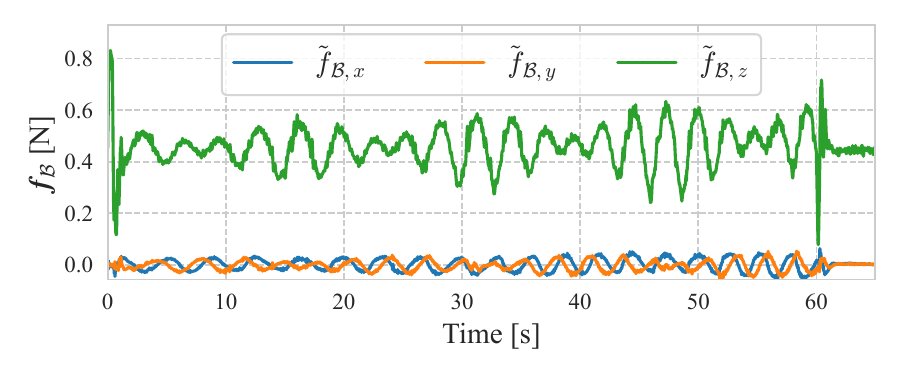}
    \caption{Example of measured forces in body frame. $f_{\mathcal{B},x}$ and $f_{\mathcal{B},y}$ are the non-ideal lateral force components during the 
    \emph{Melon} trajectory (run~1).}
    \label{fig:forces_non_idealities}
\end{figure}

\paragraph{Limitations of the physical model}
Thrust predictions obtained with the estimated $k_F$ are compared with experimental measurements of the Melon trajectory in the top panel of Figure~\ref{fig:thrust_yaw_comparison}. 
The estimated thrust along $z$ aligns closely with the measurements. However, as shown in Figure~\ref{fig:forces_non_idealities}, the reconstructed body frame forces exhibit additional components in the $x$ and $y$ directions that the model does not explain.

The reconstructed yaw torque $k_M (-\Omega_1^2 + \Omega_2^2 -\Omega_3^2 + \Omega_4^2)$ is shown together with the ground truth torque $\tilde \tau_z$ in the bottom panel of Figure~\ref{fig:thrust_yaw_comparison}. It appears that the model explains only part of the yaw dynamics.

An even larger model mismatch occurs in the roll and pitch torques, visualized in the second and third panels of Figure~\ref{fig:thrust_yaw_comparison}. For instance, the predicted pitch torque $k_F L \bigl(-\Omega_1^2 - \Omega_2^2 + \Omega_3^2 + \Omega_4^2 \bigr)$ based on the second row of~\eqref{eq:thrust_torque_mapping} departs significantly from the ground-truth $\tilde \tau_x$.

\begin{figure}
    \centering
    \includegraphics[width=\linewidth]{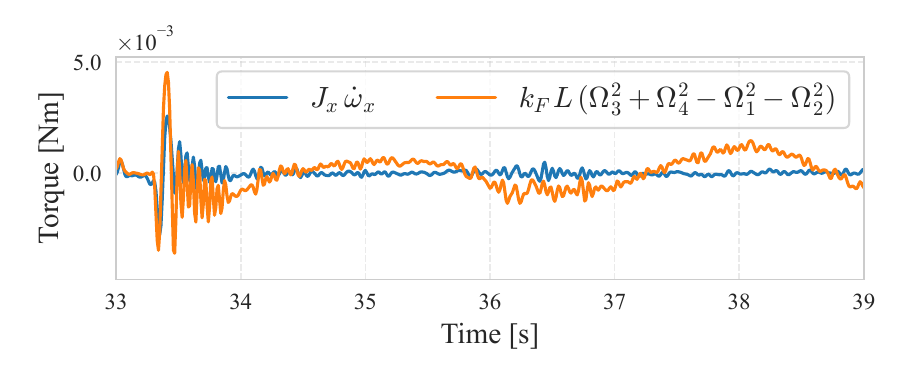}
    \caption{Zoomed comparison between measured and model-predicted thrust and yaw torque
    on the \emph{Melon} test trajectory (run1).}
    \label{fig:yaw_x_zoom}
\end{figure}

Figure~\ref{fig:yaw_x_zoom} presents a zoomed view of the ground-truth torque roll $\tilde \tau_x$, together with its prediction based on the quadratic model.
The predicted signal appears to comprise two components: a high-frequency part that agrees with the measured torque and a lower-frequency part that the model cannot account for.
This indicates that additional torques, not captured by~\eqref{eq:thrust_torque_mapping}, are acting on the drone. 

We hypothesize that the observed lack of fit arises from the limitations of the quadratic model in~\eqref{eq:thrust_torque_mapping} when applied to aggressive maneuvers, as anticipated in Remark~\ref{rem:quadratic_motor_model}.
Similar issues have been reported for larger quadrotors~\citep{bauersfeld2021neurobem, sun2019quadrotor}, although, to our knowledge, they have not yet been investigated at the nano scale.

Since this behaviour is also observed in the training data, in light of the considerations above, we adopt a simplified physical baseline in which $\dot{\boldsymbol{\omega}} = \mathbf{0}$.
We stress that this does not imply that the angular acceleration is actually zero.
Instead, we retain the portion of the physical model that accurately reflects the underlying dynamics and make a conservative modeling choice for the rotational dynamics, which we later compensate with black box components. 
Attempts to incorporate the rotational dynamics from the physical equations yielded overall worse results.

\subsection{Black-box models}

The purely data-driven models learn the state transition directly from data, without explicit physics priors.
Both architectures predict the next output as a residual correction on top of the current state.
For multi-step prediction, starting from \(\hat{\mathbf{y}}_{t \mid t} = \mathbf{\tilde{y}}_t\), the models are recursively applied as:
\begin{equation}
    \hat{\mathbf{y}}_{t+h \mid t} =
    \hat{\mathbf{y}}_{t+h-1 \mid t} +
    f_{\text{bb}}\!\big(
        \dots
    \big),
    \quad h = 1,\dots,H,
\end{equation}
where the structure of \(f_{\text{bb}}\) depends on the specific architecture described below.

\paragraph{Rolling Feed-Forward}
The rolling feed-forward (FF) model implements \(f_{\text{bb}}\) as a multilayer perceptron (MLP) that outputs the residual increment \(\Delta\hat{\mathbf{y}}\) based solely on the immediate state and input. The recursion becomes:
\begin{equation}
    \hat{\mathbf{y}}_{t+h \mid t} =
    \hat{\mathbf{y}}_{t+h-1 \mid t} +
    f_{\text{MLP}}\!\big(
        \hat{\mathbf{y}}_{t+h-1 \mid t},
        \mathbf{u}_{t+h-1}
    \big).
\end{equation}
It captures instantaneous nonlinear relationships between state and input, providing a purely static approximation of the residual dynamics.

\paragraph{Long Short-Term Memory}
The LSTM model extends the residual formulation by introducing a recurrent hidden state \(\mathbf{h}\), which evolves dynamically over the prediction horizon.
To properly initialize the internal dynamics from the static observation \(\mathbf{y}_t\), the initial hidden state is inferred via a dedicated mapping \(\mathbf{h}_{t \mid t} = f_{\text{init}}(\mathbf{\tilde{y}}_t)\), parameterized as a small MLP.
The multi-step prediction is then generated via the coupled update law:
\begin{subequations}
\begin{align}
    \mathbf{h}_{t+h \mid t} &= \phi_{\text{LSTM}}\big(\mathbf{h}_{t+h-1 \mid t}, \hat{\mathbf{y}}_{t+h-1 \mid t}, \mathbf{u}_{t+h-1}\big), \\[3pt]
    \hat{\mathbf{y}}_{t+h \mid t} &= \hat{\mathbf{y}}_{t+h-1 \mid t} + \psi_{\text{out}}\big(\mathbf{h}_{t+h \mid t}\big),
\end{align}
\end{subequations}
for \(h = 1,\dots,H\).
Here, \(f_{\text{init}}\) projects the initial system state into the latent space, \(\phi_{\text{LSTM}}\) propagates the temporal context, and \(\psi_{\text{out}}\) is a linear readout layer mapping the updated hidden state to the residual increment.

\subsection{Hybrid model}
The hybrid model (Physical + Residual) extends the physical predictor \(f_{\text{phys}}\) with a neural correction term that learns to compensate for unmodeled residual effects.
Starting from the current measurement \(\mathbf{y}_t\), the model recursively predicts future outputs as:
\begin{equation}
\begin{aligned}
    \hat{\mathbf{y}}_{t \mid t} &= \mathbf{\tilde{y}}_t, \\[3pt]
    \hat{\mathbf{y}}_{t+h \mid t} &=
    f_{\text{phys}}\!\big(
        \hat{\mathbf{y}}_{t+h-1 \mid t},
        \mathbf{u}_{t+h-1}
    \big)
    + {}\\[-3pt]
    &\quad f_{\text{res}}\!\big(
        \hat{\mathbf{y}}_{t+h-1 \mid t},
        \mathbf{u}_{t+h-1}
    \big),
    \quad h = 1,\dots,H.
\end{aligned}
\label{eq:residual_model}
\end{equation}
where \(f_{\text{phys}}\) denotes the discrete-time physical mapping defined in~\eqref{eq:phys_discrete}, and \(f_{\text{res}}\) is a feed-forward neural network that outputs a residual correction \(\Delta\mathbf{y}_t\) in the same representation as \(\mathbf{y}\).
The residual component is initialized to zero and progressively learns to correct systematic errors of the physical model, such as aerodynamic drag, blade interference, or motor lag.

\subsection{Performance discussion}

\begin{figure*}[t]
    \centering
    \includegraphics[width=\linewidth]{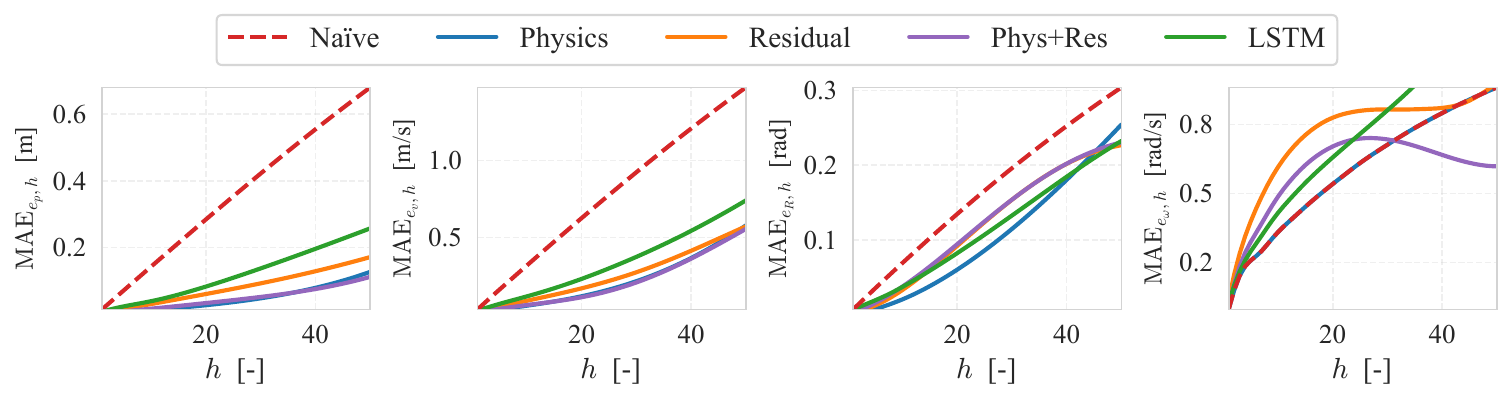}
    \caption{$\mathrm{MAE}_{(\cdot),h}$ scores (for position, linear velocity, orientation and angular velocity) for $h=1, \ldots,50$}
    \label{fig:metrics_models}
\end{figure*}

\begin{table*}[t]
    \centering
    \caption{Numerical performance at $h=1,10,50$. The italic column reports the cumulative simulation error (sum of MAEs over $h=1..50$).}
    \setlength{\tabcolsep}{3pt}
    \scriptsize
    \renewcommand{\arraystretch}{1.2}
    \begin{tabular}{l|cccc|cccc|cccc|cccc}
        \toprule
        & \multicolumn{4}{c|}{$\mathrm{MAE}_{p,h}$ [m]}
        & \multicolumn{4}{c|}{$\mathrm{MAE}_{v,h}$ [m/s]}
        & \multicolumn{4}{c|}{$\mathrm{MAE}_{R,h}$ [rad]}
        & \multicolumn{4}{c}{$\mathrm{MAE}_{\omega,h}$ [rad/s]}\\[1mm]

        Model
        & $h{=}1$ & $h{=}10$ & $h{=}50$ & \textit{$h=1{:}50$}
        & $h{=}1$ & $h{=}10$ & $h{=}50$ & \textit{$h{=}1{:}50$}
        & $h{=}1$ & $h{=}10$ & $h{=}50$ & \textit{$h{=}1{:}50$}
        & $h{=}1$ & $h{=}10$ & $h{=}50$ & \textit{$h{=}1{:}50$}\\
        \midrule
\toprule
\midrule
Na\"ive & 0.0143 & 0.1430 & 0.6797 & \textit{17.7878} & 0.0329 & 0.3182 & 1.4749 & \textit{38.9241} & 0.0071 & 0.0692 & 0.3041 & \textit{8.2138} & \textbf{0.0796} & \textbf{0.3596} & 0.8837 & \textit{29.0866} \\
Physics & \textbf{0.0013} & \textbf{0.0126} & 0.1269 & \textbf{\textit{2.3223}} & \textbf{0.0080} & \textbf{0.0570} & 0.5781 & \textit{10.6232} & \textbf{0.0011} & \textbf{0.0205} & 0.2544 & \textbf{\textit{5.1013}} & \textbf{0.0796} & \textbf{0.3596} & 0.8837 & \textit{29.0866} \\
Residual & 0.0032 & 0.0305 & 0.1712 & \textit{4.0519} & 0.0116 & 0.0890 & 0.5720 & \textit{12.5809} & 0.0022 & 0.0331 & \textbf{0.2268} & \textit{6.0591} & 0.1138 & 0.5949 & 0.9005 & \textit{35.5735} \\
Phys+Res & 0.0016 & 0.0166 & \textbf{0.1119} & \textit{2.3625} & 0.0092 & 0.0613 & \textbf{0.5556} & \textbf{\textit{10.4033}} & 0.0027 & 0.0376 & 0.2306 & \textit{6.1534} & 0.0912 & 0.4880 & \textbf{0.5979} & \textbf{\textit{28.9873}} \\
LSTM & 0.0079 & 0.0390 & 0.2572 & \textit{5.9711} & 0.0247 & 0.1175 & 0.7407 & \textit{16.7207} & 0.0066 & 0.0372 & 0.2325 & \textit{5.6525} & 0.1021 & 0.4292 & 1.2407 & \textit{35.8353} \\
\bottomrule
\bottomrule
\end{tabular}
\label{tab:numerical_performance}
\end{table*}

\begin{figure*}[!t]
    \centering
    \includegraphics[width=\linewidth]{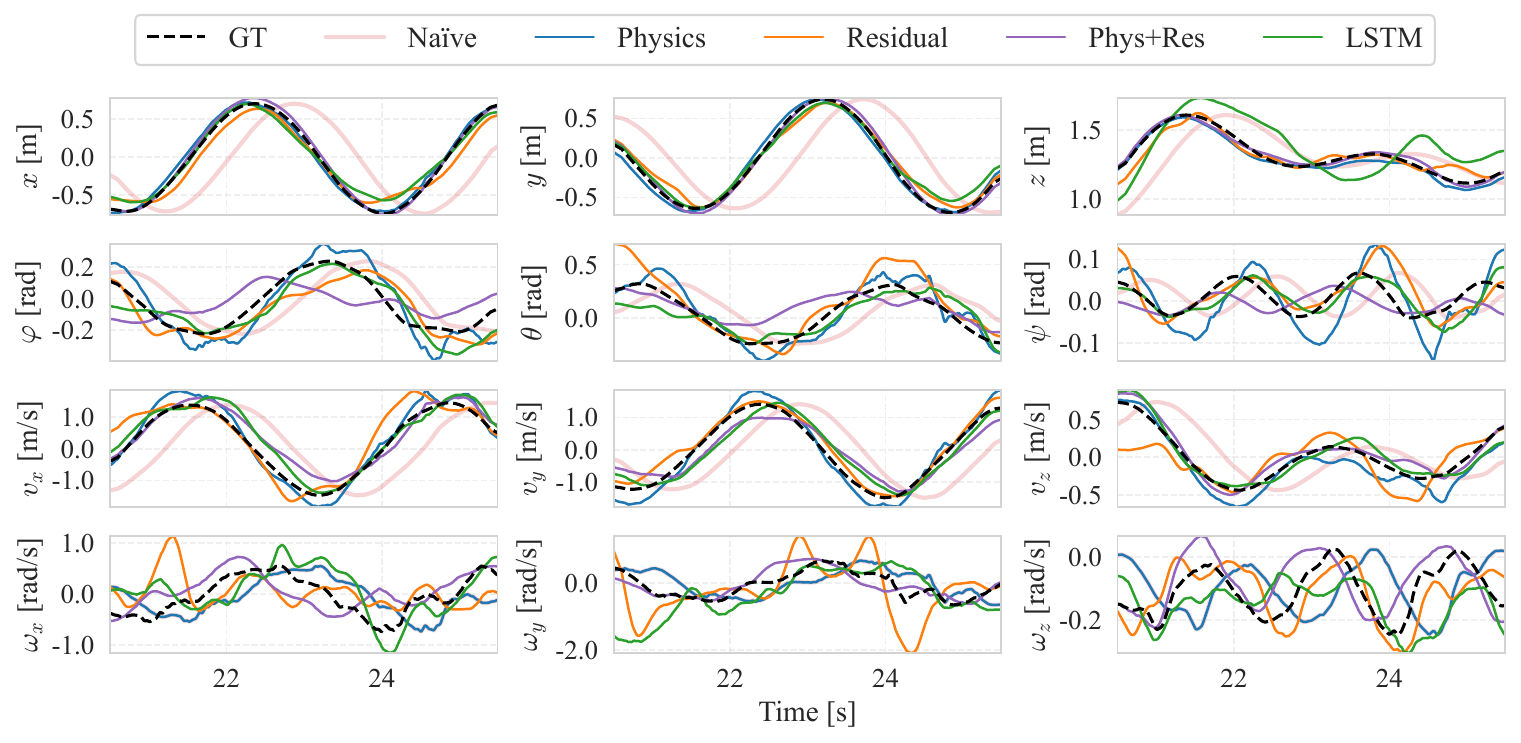}
    \caption{Output trajectories of \textit{Melon} (run 1), from 20 to 25 s, comparing 50-step ahead predictions $\mathbf{\hat{y}}_{t+50|t}$ from different models with respect the ground truth $\mathbf{y}_{t+50}$ Trajectories have been filtered (centered rolling mean with window 20) for visualization purpose.}
    \label{fig:lineplots_models}
\end{figure*}

Figure~\ref{fig:metrics_models} reports the full $\mathrm{MAE}_{(\cdot),h}$ curves for $h=1,\ldots,50$ across all models and averaged across the three runs of the \textit{Melon} trajectory, while Figure~\ref{fig:lineplots_models} illustrates qualitatively the 50-step-ahead predictions on a representative segment of the first test trajectory. For clarity, Table~\ref{tab:numerical_performance} summarizes the MAE at three representative horizons ($h=1$, $h=10$, and $h=50$) for the five models considered: \emph{Na\"ive}, \emph{Physics}, \emph{Residual}, \emph{Phys.\,+\,Res.}, and \emph{LSTM}.

All learned models operate in a \emph{residual} fashion: they predict state increments and construct trajectories through recursive summation, instead of predicting absolute states.
Formally, each model learns:
\begin{equation}\label{eq:residuals}
    \boldsymbol{\Delta \hat{y}}_{t+h-1\mid t}
    = \boldsymbol{\hat{y}}_{t+h\mid t}
    - \boldsymbol{\hat{y}}_{t+h-1\mid t},
\end{equation}
and the multi-step prediction is obtained by accumulating these residuals. 
Note that any systematic bias in the learned increments compounds over time, making long-horizon prediction sensitive to model structure.
Models without strong inductive priors -- notably the pure neural residual model and the LSTM -- are therefore more prone to drift.
Indeed, feed-forward neural networks must approximate the full transition map directly from noisy data, making them especially sensitive to bias in the learned residual increments.
The LSTM faces an additional limitation: its recurrent update is governed by an internal hidden state rather than explicit physical variables, making it challenging to enforce physical consistency and causing hidden-state drift to accumulate over time.
In contrast, the Phys+Res model benefits from the stability of the physical integrator, requiring the network to learn only minor corrections to account for unmodelled effects.
This structural prior provides a more reliable foundation for multi-step prediction, explaining its superior long-horizon performance.

For position and linear velocity, all learned models outperform the na\"ive integrator.
The physical model and the physics--plus--residual network deliver the best performance, followed by the purely neural residual model and the LSTM.
Translational dynamics are comparatively easier to identify: the mass is known with high accuracy, and the linear accelerometer provides low-noise measurements, making the thrust contribution to linear acceleration very reliable, even if torque estimation is more uncertain.
This behaviour is clearly visible in the first row of Figure~\ref{fig:lineplots_models}: the 50--step--ahead predictions from the learned models align closely with the \emph{ground truth} (GT), in sharp
contrast with the Na\"ive prediction (pink), which behaves as a pure delay. 
An analogous pattern is observed for the linear velocities (third row): despite the higher variability inherent to these signals, the learned models remain temporally aligned with the GT, whereas the Na\"ive fails to track the correct timing of the motion.

Rotational dynamics are particularly challenging to predict due to limitations in the standard quadratic motor model, which does not capture a slow, low-frequency component of the actuation dynamics and therefore introduces systematic errors in the torque prediction.
The physics model mitigates this by driving the rotational motion directly from the measured angular velocity and setting its derivative to zero, preventing the instability that would arise from integrating an incorrect torque model and explaining its comparatively strong performance at short horizons.
Learned residual models, both the feedforward and the LSTM, do not substantially improve rotational accuracy, as they are trained and evaluated on 50-step windows that are too short to resolve the underlying slow dynamics.
This choice is deliberate: the benchmark targets 0.5-s multi-step prediction, and longer temporal windows would be incompatible with real-time deployment on embedded hardware such as the Crazyflie’s STM32 microcontroller. 
Moreover, even if trained on longer sequences, test-time rollouts would still be limited to the same short horizons, preventing reconstruction of long-timescale actuation effects.
As a result, improvements in rotational prediction remain modest across baselines, pointing to the need for either more expressive temporal models with longer context or enhanced physics-based descriptions of rotor and motor dynamics.

\begin{table}[t]
    \centering
    \setlength{\tabcolsep}{6pt}
    \renewcommand{\arraystretch}{1.2}
    \caption{\#$^{\mathrm{params}}$ indicates the number of trainable parameters, Compute denotes the floating-point operations (FLOPs) per step, and $T^{\mathrm{inf}}$ is the average inference time measured on the target STM32 MCU.}
    \resizebox{\linewidth}{!}{%
    \begin{tabular}{l|ccc}
        \toprule
        Model
        & {\#$^{\mathrm{params}}$ [k]} & {Compute [kFLOPs]} & {$T^{\mathrm{inf}}$ [ms]}  \\
        \midrule
        Na\"ive &  {--} & {--} &   {--} \\
        Physics  &   {--} & 0.8 & 1.79 \\
        Residual   &  18.5 & 18.2 & $\textbf{1.03}$ \\
        Phys+Res &  18.5 & 18.2 & 2.82 \\
        LSTM     &  24.7 & 25.1 & 2.09 \\
        \bottomrule
    \end{tabular}
    }
    \label{tab:models_params}
\end{table}

Table~\ref{tab:models_params} summarizes the main architectural and computational characteristics of the evaluated models, which we encourage practitioners to take into account when designing or assessing new benchmarks.
In particular, the table lists the number of trainable parameters for each model.
All learned architectures were selected to have a comparable order of magnitude in terms of parameter count, thereby avoiding biased comparisons and, most importantly, preventing excessively demanding computational loads on embedded hardware.
The second column reports the number of floating-point operations (FLOPs) per forward pass.
We further deploy all models in 32-bit float precision on our platform's STM32 MCU with STM32EdgeAI Developer Cloud\footnote{\url{https://stedgeai-dc.st.com}. Model export process and benchmark setup are described in our GitHub repo.}.
This service enables model benchmarking on the target MCU without on having access to a Crazyflie drone, leveraging the service's test farm of physical MCUs in the cloud.
We measure the resulting inference time $T^{\mathrm{inf}}$, reflecting the computational cost of running each model \emph{on board}.
Among the learned architectures, the LSTM is the most computationally demanding, with an inference time of $2.09\,\mathrm{ms}$ for a single one-step prediction.
This is approximately twice the inference time of the purely residual feed-forward model, which is the lightest architecture. 

For $N$-step--ahead predictions, the computational cost scales linearly as $N \times T^{\mathrm{inf}}$. At a sampling rate of $100\,\mathrm{Hz}$, this implies that only $4$ to $8$ steps could be predicted within a single control cycle, which may still be suitable for a short-horizon model predictive control.
It is worth noting that the Physics model (and, consequently, the Phys.\,+\,Res.\ model) is slower than the purely residual network despite having zero trainable parameters. 
This is due to the more expensive operations inherent to rigid-body dynamics, such as cross products, normalization steps, and matrix--vector computations, which impose a higher computational burden on embedded hardware.
\section{Conclusions}\label{sec:conclusions}

We have presented a comprehensive benchmark for nonlinear system identification on nano-quadrotors, comprising high-frequency flight data, a standardized evaluation protocol, and a suite of baseline models.
Our comparative analysis demonstrates that while current methods achieve high-fidelity prediction of position and linear velocity, making them viable for model-based control, accurate modeling of rotational dynamics remains a significant open challenge.
A primary bottleneck is the standard quadratic motor model, which neglects low-frequency actuation dynamics that residual feedforward corrections alone cannot fully compensate. 
This finding suggests that future hybrid approaches must leverage temporal learning architectures or incorporate more expressive physics-based rotor models. 
Finally, while the proposed trajectories are sufficiently aggressive to excite nonlinearities, improved actuation modeling will likely unlock the platform’s full flight envelope. 
This creates a virtuous cycle where better models enable more agile data collection, further enriching the benchmark and accelerating progress in control for miniaturized aerial vehicles.

\section*{Acknowledgments}
Supported by IDSIA.

\bibliographystyle{unsrtnat} 
\bibliography{cas-refs}

\end{document}